\documentclass[conference,compsoc]{IEEEtran}

\let\OLDthebibliography\thebibliography
\renewcommand\thebibliography[1]{
  \OLDthebibliography{#1}
  \setlength{\parskip}{0pt}
  \setlength{\itemsep}{0pt plus 0.3ex}
}

\usepackage[T1]{fontenc}
\usepackage{soul}
\usepackage{comment}
\usepackage{todonotes}
\usepackage{array}
\usepackage{amsmath}
\usepackage{amssymb}
\usepackage{cite}
\usepackage{graphicx} 
\usepackage{xspace}
\usepackage{url}
\usepackage[hidelinks]{hyperref}
\usepackage[capitalise]{cleveref}

\usepackage[linesnumbered,ruled,vlined,algo2e]{algorithm2e}
\usepackage{color}
\usepackage{subcaption}
\usepackage{lipsum}
\usepackage{footmisc}
\usepackage[group-separator={,},group-minimum-digits=4]{siunitx}
\usepackage{comment}
\usepackage{enumitem}
\usepackage{makecell}
\usepackage{colortbl}
\usepackage{dsfont}

\usepackage{multirow}
\usepackage{arydshln}

\usepackage{cuted}

\usepackage[dvipsnames]{xcolor}

\usepackage[section]{placeins}

\usepackage{booktabs}

\definecolor{mycitecolor}{rgb}{0.5,0,0}
\definecolor{mylinkcolor}{rgb}{0, 0.2, 1.0}
\definecolor{myurlcolor}{rgb}{0.0, 0.28, 0.67}

\newcommand{\etal}{\textit{et al.}}

\definecolor{turquoise}{rgb}{0.19, 0.84, 0.78}
\definecolor{burgundy}{rgb}{0.5, 0.0, 0.13}
\definecolor{burntorange}{rgb}{0.8, 0.33, 0.0}
\definecolor{deepcarrotorange}{rgb}{0.91, 0.41, 0.17}
\definecolor{darkgray}{rgb}{0.66, 0.66, 0.66}
\definecolor{lightgray}{rgb}{0.77, 0.77, 0.77}
\definecolor{myGreen}{rgb}{0,0.51,0}

\SetCommentSty{mycommfont}
\SetKwInput{KwInput}{Input}                
\SetKwInput{KwOutput}{Output}              

\newcommand{\code}[1]{{\small \texttt{#1}}}

\mathchardef\mhyphen="2D 
\newcommand{\subheading}[1]{\vspace{0.5 em}\noindent \textbf{#1}}

\newcommand{\curtime}{\ensuremath{t}\xspace}
\newcommand{\graph}{\ensuremath{G}\xspace}

\newcommand{\graphind}[1]{\graph_{#1}\xspace} 

\newcommand{\networksize}{\ensuremath{n}\xspace}
\newcommand{\corrsize}{\ensuremath{\networksize_{cor}}\xspace} 
\newcommand{\isosize}{\ensuremath{\networksize_{iso}}\xspace} 
\newcommand{\crashsize}{\ensuremath{\networksize_{cr}}\xspace} 

\newcommand{\nodeset}{\ensuremath{\mathcal{N}}\xspace}
\newcommand{\nodesetind}[1]{\nodeset_{#1}\xspace}
\newcommand{\nodesetindext}[2]{\nodeset_{#1}^{#2}\xspace}
\newcommand{\node}{\ensuremath{N}\xspace}

\newcommand{\edgeset}{\ensuremath{\mathcal{E}}\xspace}
\newcommand{\edgesetind}[1]{\edgeset_{#1}\xspace}

\newcommand{\ERconnectivity}{\ensuremath{p}\xspace}
\newcommand{\nodedegree}{\ensuremath{d}\xspace}

\renewcommand{\subheading}[1]{\vspace{0.25 em}\noindent \textbf{#1}}

\newcommand{\setsize}[1]{\ensuremath{\left|#1\right|}\xspace}

\newcommand{\abs}[1]{\ensuremath{\left|#1\right|}\xspace}

\newcommand{\peers}[1]{\ensuremath{\mathsf{Peers}(#1)}}

\newcommand{\nameprotocol}{gossip protocol\xspace}
\newcommand{\bigcapnameprotocol}{Gossip Protocol\xspace}
\newcommand{\nameprotocols}{gossip protocols\xspace}

\newcommand{\protocol}{\ensuremath{\pi}\xspace}

\newcommand{\stateset}{\ensuremath{\mathcal{S}}\xspace}
\newcommand{\state}{\ensuremath{s}\xspace}

\newcommand{\stateInit}{\ensuremath{\state_{\mathsf{init}}}\xspace}

\newcommand{\stateFinal}{\ensuremath{\state_{\mathsf{final}}}\xspace}

\newcommand{\nodestate}[1]{\ensuremath{\mathsf{State}({#1})}\xspace}

\newcommand{\starttime}{\ensuremath{\curtime_{\mathsf{start}}}\xspace}
\newcommand{\completedtime}{\ensuremath{\curtime_{\mathsf{end}}}\xspace}

\newcommand{\maxdelay}{\ensuremath{\Delta}\xspace}

\newcommand{\maxcoverage}{\ensuremath{\Omega}\xspace}

\newcommand{\robust}{\ensuremath{\mathsf{Quality}}\xspace}
\newcommand{\namerobust}{quality\xspace}

\newcommand{\attacker}{\ensuremath{\mathcal{A}}\xspace}
\newcommand{\corrupt}{\ensuremath{\mathsf{Corrupt}}\xspace}
\newcommand{\corruptdelay}{\ensuremath{\delta_{cor}}\xspace}
\newcommand{\corruptfrac}{\ensuremath{\beta}\xspace}
\newcommand{\bound}{\ensuremath{\corruptfrac}\xspace}
\newcommand{\corruptednodesetind}[1]{\ensuremath{\nodesetindext{#1}{cor}}}
\newcommand{\honestnodesetind}[1]{\ensuremath{\nodesetindext{#1}{hon.}}}

\newcommand{\arrivalrate}{\ensuremath{\lambda}\xspace}
\newcommand{\droprate}{\ensuremath{\xi}\xspace}

\newcommand{\churnrate}{\ensuremath{c}\xspace}

\newcommand{\bftboundtext}{BFT quorum size\xspace}
\newcommand{\nsboundtext}{NS quorum size\xspace}

\newcommand{\bftrelbound}{\ensuremath{1/3\cdot(1-1/n)}\xspace}

\begin{document}

\title{On the Robustness of Gossip Protocols in Modern Blockchains}

\makeatletter
\newcommand{\linebreakand}{%
  \end{@IEEEauthorhalign}
  \hfill\mbox{}\par
  \mbox{}\hfill\begin{@IEEEauthorhalign}
}
\makeatother

\author{
    \IEEEauthorblockN{
        Wenting Li\IEEEauthorrefmark{1},
        Jannik Albrecht\IEEEauthorrefmark{1},
        Frederik Armknecht\IEEEauthorrefmark{3}, and
        Ghassan Karame\IEEEauthorrefmark{1}
    }
    \linebreakand
    \IEEEauthorblockA{
        \IEEEauthorrefmark{1}Ruhr University Bochum, Germany\\ 
        first.last@ruhr-uni-bochum.de
    }
    \and
    \IEEEauthorblockA{
        \IEEEauthorrefmark{3}University of Mannheim, Germany\\
        armknecht@uni-mannheim.de
    }
}

\maketitle

\begin{abstract}
Modern platforms such as Ethereum and Polkadot nowadays execute Byzantine fault-tolerant consensus protocols over large validator committees, often comprising hundreds of nodes. At this scale, complete graph topologies for connecting nodes are impractical, leading systems to adopt various gossip mechanisms, often in an ad hoc manner.

In this paper, we study the robustness of widely deployed gossip protocols used in committee-based blockchains against realistic network adversaries. We show, both analytically and through extensive empirical measurements, that the graph parameters commonly chosen to instantiate random graphs are insufficient to guarantee that enough honest nodes receive consensus messages for quorum formation in existing protocols. Motivated by this finding, we empirically assess the robustness of Bitcoin's Erd\H{o}s--R\'enyi (ER) graphs, Polkadot’s grid topology, Ethereum’s GossipSub, and Solana’s chunk- and tree-based disseminations. Our results reveal surprising tradeoffs: Solana’s dynamic trees are bandwidth-efficient in benign settings but fragile under adaptive attacks; Polkadot’s grid achieves strong robustness at the cost of high bandwidth overhead; GossipSub is both bandwidth-efficient and resilient to adaptive adversaries, yet highly sensitive to parameter choices; and ER graphs provide a strong robustness–efficiency tradeoff, e.g., when the (average) fanout degree exceeds 15 for networks up to 5000 nodes.
\end{abstract}

\IEEEpeerreviewmaketitle

\section{Introduction}\label{sec:intro}

Decentralized platforms are rapidly gaining ground as a key technology in blockchain deployments, machine learning (e.g., federated and decentralized learning~\cite{DBLP:conf/aistats/McMahanMRHA17,DBLP:conf/nips/LianZZHZL17,DBLP:conf/icml/TangLYZL18}), social networks (e.g., federated social networks such as Mastodon~\cite{mastodon,mastodon_code}), file storage (e.g., IFPS~\cite{ipfs}), and content and music distribution (e.g., Odysee~\cite{odysee} and Audius~\cite{audius}).
At the core of almost all decentralized deployments lies a peer-to-peer (P2P) network that is responsible for reliably propagating critical information and metadata to all nodes. For instance, the security analysis of blockchain protocols often assumes a synchronous model where messages arrive at all nodes within a bounded delay window~\cite{tempering-delivery}. Any partition in the P2P network, however, can break these guarantees~\cite{albrecht2024larger}, potentially causing major forks~\cite{DBLP:journals/ieeesp/GervaisKCC14}. 

The interplay between the network layer and the consensus layer has been largely overlooked, with most existing designs treating these two components almost in isolation. This separation appears to be an artifact of the field's development.
Historically, BFT protocols were deployed on only a small number of nodes. With such a limited scale, a complete graph was the natural choice: every node could communicate directly with every other, keeping overhead tolerable.
Today, however, platforms such as Ethereum, Polkadot, and Cosmos run BFT-style consensus across large validator committees. For the first time in decades, BFT protocols operate at scale, involving hundreds or even thousands of participants—for example, Ethereum and Polkadot maintain validator committees of roughly \num{9000} and \num{600} nodes, respectively.
This substantial increase in committee size renders the reliance on complete graphs impractical. Consequently, modern decentralized platforms have replaced them with more sophisticated dissemination protocols, adapted from the P2P networking literature, to enable validators to exchange information efficiently and reach consensus at scale. Ethereum, for example, uses GossipSub~\cite{gossipsub} to organize nodes into a mesh network in which full messages are pushed across peers, complemented by metadata-only links that periodically advertise information about recently seen messages. Polkadot, in contrast, uses a grid-like dissemination protocol that guarantees a maximum two-hop distance between any two nodes, ensuring rapid and efficient information propagation.

This disconnect (i.e., designing network protocols and consensus protocols in isolation) is not only inefficient but also affects safety and liveness. For example, consider the graphs in Figure~\ref{fig:graph_examples}, each consisting of $\networksize=7$ nodes, where nodes run a classical BFT-style consensus algorithm (similar to that used in Ethereum) to agree on the next block. Here, an adversary may corrupt up to two nodes, which falls well within the BFT compromise threshold of $\leq\frac{1}{3}(\networksize-1)$.
In Figure~\ref{fig:separable_graph_2}, corrupting nodes $c$ and $d$ suffices to reduce the largest connected component of honest nodes to at most two nodes. Since coverage (i.e., the maximum number of honest nodes that receive a gossip) is then too limited to reach the minimum quorum size of $\frac{1}{3}(2\networksize+1)=5$ required for classical BFT consensus, the protocol is likely to fail in making progress. The graph in Figure~\ref{fig:robust_graph}, by contrast, tolerates the corruption of any two nodes without the coverage dropping below $5$. In other words, the graph in Figure~\ref{fig:robust_graph} is clearly \emph{more robust} than that of Figure~\ref{fig:separable_graph_2}.

In this paper, we address the fundamental problem of evaluating the robustness of gossip protocols under \emph{realistic} adversarial strategies. 
While the examples in Figure~\ref{fig:graph_examples} are relatively straightforward, the challenge is that consensus security is well established in theoretical models, whereas gossip effectiveness is typically measured through empirical metrics such as coverage and delay. Bridging the two requires a unifying model that captures the security-relevant dynamics of churn and message dissemination, together with metrics that translate gossip behavior into security properties under a realistic adversary.
Our analysis aims to answer the following research questions:
\begin{description}[leftmargin=*]
    \item[RQ1:] Under which parameters and settings are gossip protocols robust against realistic adaptive adversaries?
    \item[RQ2:] How do various gossip protocols compare with respect to their robustness and performance?
\end{description}

We address these questions as follows. First, we introduce a novel framework that captures both the robustness and the efficiency of existing gossip protocols once peer discovery has completed and the overlay topology has formed (Section~\ref{sec:model}). The framework incorporates a practical threat model with targeted attacks and \emph{mobile} adversaries, which can release corrupted nodes and corrupt others as the network evolves. Such adversaries adapt their choice of corrupted nodes based on \emph{readily available} information about node placement within the network. Leveraging our framework, we analyze the robustness of random graph topologies and show, both analytically and empirically, that the default parameters used in currently deployed random graph topologies (such as Bitcoin's fanout degree of $8$, or the de facto standard fanout degree of $\ln{(\networksize)}$ in random graphs of size $\networksize$~\cite{Erdos_Renyi}) are ill-suited for composition with classical Byzantine Fault Tolerance (BFT) protocols (Section~\ref{sec:security-model}).

\begin{figure}
    \centering
    \begin{subfigure}[c]{0.47\linewidth}
        \includegraphics[width=\textwidth]{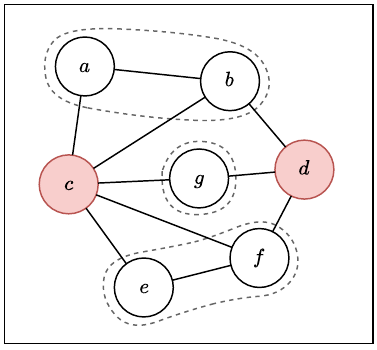}
        \caption{Example 1}
        \label{fig:separable_graph_2}
    \end{subfigure}
    \begin{subfigure}[c]{0.47\linewidth}
        \includegraphics[width=\textwidth]{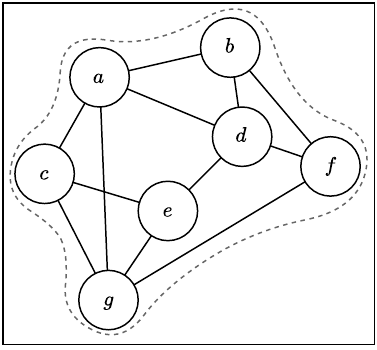}
        \caption{Example 2}
        \label{fig:robust_graph}
    \end{subfigure}
    \caption{Two examples with different graph topologies. A compromise of two nodes in Figure~\ref{fig:separable_graph_2} reduces the coverage to $2$ and hence prevents the formation of a quorum, whereas a compromise of any two nodes in the more robust graph in Figure~\ref{fig:robust_graph} does not affect gossip reception by the remaining honest nodes. Dashed lines highlight connected components. } 
    \label{fig:graph_examples}
\end{figure}

We then use this observation as a guiding example and extend our analysis to modern gossip protocols deployed in real-world committee-based blockchains, including Polkadot’s grid-based gossip~\cite{polkadot-grid}, Ethereum’s GossipSub~\cite{gossipsub}, and Solana’s chunk- and tree-based protocols~\cite{solana-turbine}. 
More specifically, we considerably extend the popular blockchain network simulator SimBlock~\cite{simblock} to measure delay (the maximum time for a gossip to spread) and coverage (maximum number of nodes that receive the gossip) of the gossip mechanisms employed by Polkadot, Ethereum, and Solana with respect to various parameters, such as the network size, the fanout degree, the churn rate, and various adversarial strategies (Section~\ref{sec:evaluation}).
Our analysis in Section~\ref{sec:findings} reveals the following surprising results: 
\begin{itemize}[leftmargin=*]
    \item In the absence of attacks, Solana’s gossip protocol (i.e., dynamic trees combined with chunk-based dissemination) achieves coverage comparable to that of other gossip protocols while incurring the lowest bandwidth overhead. However, under realistic adaptive attacks, Solana exhibits the lowest robustness and fails to guarantee the minimum coverage required to form a quorum.
    \item Polkadot’s grid topology, in contrast, maintains consistently high robustness even in the presence of mobile adversaries, but this robustness comes at the cost of substantial bandwidth overhead, exceeding that of Solana by more than a factor of $20$. 
    \item GossipSub achieves moderate to high coverage under mobile adversaries and is the most bandwidth-efficient protocol in this setting, as it actively excludes poorly performing peers from dissemination paths. This efficiency, however, depends critically on careful parameter tuning: if the expected message rate or network latency is misestimated, GossipSub may disconnect not only corrupted peers but also honest ones, leading to a rapid degradation in robustness. 
    \item Random graphs fail to provide sufficient robustness against mobile/adaptive attacks when configured with a fanout of 8, as in Bitcoin.\footnote{Coincidentally, with an estimated network size of $\networksize=5,000$ in Bitcoin, this closely matches the Erd\H os-R\'enyi threshold~\cite{Erdos_Renyi}, where connectivity is almost guaranteed if the average degree is $\approx\ln{(\networksize)}$.}  However, with an appropriately chosen fanout degree ($\geq 15 \gg \ln{(\num{5000})}$ for up to $\networksize=\num{5000}$ nodes), they strike a favorable trade-off between robustness and bandwidth efficiency, in some cases exceeding the provisions of more sophisticated protocols.  
    \item Last but not least, network churn significantly further degrades the robustness of existing gossip protocols, preventing quorum formation across most studied protocols. 
\end{itemize}
\section{System \& Network Model} \label{sec:model}
In this section, we introduce our model for capturing the process of data dissemination in (potentially dynamic) communication networks. 
Our model extends existing graph models along two dimensions: it incorporates dynamic topologies that account for churn and introduces node states to represent information dissemination. The extended graph model not only forms the basis for the evaluation in Section~\ref{sec:evaluation}  but also allows for characterizing new attacks that yield the bounds in Section~\ref{sec:Breaking-BFT-Bound}.

\subsection{Communication Networks} \label{sec:network}
We model a communication network as an undirected graph $\graph=(\nodeset,\edgeset)$, where nodes $ \nodeset$ represent the members of the network, and the edges $\edgeset\subseteq \nodeset\times\nodeset$ their communication links. Permissionless blockchain networks are commonly modeled by Erd\H{o}s--R\'enyi graphs  $G(\networksize,\ERconnectivity)$. These are randomly connected graphs with \networksize nodes, where each edge between any two nodes exists independently with probability \ERconnectivity, resulting in an average node degree of $\nodedegree=\ERconnectivity\cdot (\networksize-1)$. 

The peers $\peers{\node}$ of a node \node are all neighboring nodes, i.e., nodes directly connected to \node by an edge. To evaluate a \nameprotocol, it is important to check if a gossip spawned at some node $\node$ eventually reaches another node $\node'$. A pair of two distinct nodes $(\node,\node')$ is called \emph{connected} if a path exists from $\node$ to $\node'$. Consequently, a graph \graph is connected if any pair of distinct nodes is connected. 
The largest connected component of a graph $\graph=(\nodeset,\edgeset)$ is the largest subgraph $\graph'=(\nodeset',\edgeset')$ with $\nodeset'\subseteq\nodeset$ and $\edgeset'\subseteq\edgeset$ such that all pairs of nodes in $\nodeset'$ are connected.

In practice, graph topology may change over time due to churn. To keep track of these changes, we index the graphs with a discrete-time counter \curtime: $\graphind{\curtime}=(\nodesetind{\curtime},\edgesetind{\curtime})$.\footnote{When the time \curtime is clear from the context or the graph does not change for the period considered, we may omit the parameter \curtime.}
We capture churn via the churn rate \churnrate, which represents the fraction of nodes that may leave and rejoin the network, with each node's online and offline durations drawn from a specified distribution.

\subsection{\bigcapnameprotocol}\label{sec:protocol}
 
The goal of a \nameprotocol is to disseminate data across the network as thoroughly and rapidly as possible. To evaluate the effectiveness of such a protocol, we must track how quickly information spreads among the nodes. To this end, we model each node as being stateful. 

We denote by \stateset the set of all possible states of nodes and by $\nodestate{\node,\curtime}\in\stateset$ the state of node \node at point in time \curtime. Moreover, we denote by  $\nodesetind{\curtime,\state}$ the set of all benign nodes in $\nodesetind{\curtime}$ that are in state \state at time \curtime. \stateset contains two special states: \stateInit and \stateFinal. The \emph{initial} state \stateInit characterizes nodes that are completely unaware of the data being disseminated. In contrast, a node is in the \emph{final state} $\stateFinal$ if it knows the full content of that data. 

When the \nameprotocol starts at time $\starttime$, typically $\node^*$ marks the origin of the gossip to be disseminated, while all other nodes are not informed (yet). Formally, this means that $\nodesetind{\starttime,\stateFinal}= \{\node^*\}$ and $\nodesetind{\starttime,\stateInit}= \nodesetind{\starttime}\setminus \{\node^*\}$. Then, at subsequent epochs $\curtime>\starttime$, the following two actions take place at each node $\node$. 
First, $\node$ updates its state from $\nodestate{\node,\curtime-1}$ to $\nodestate{\node,\curtime}$, based on the messages it received in the previous epoch $\curtime-1$. Then, it sends some messages to its peers $\peers{\node}$, depending on its state $\nodestate{\node,\curtime}$. Here, we assume that each node \node knows which of its peers are in a final state already. If this holds for all peers,   \node will stop sending any messages in this protocol run, but still may receive messages from other nodes if \node is not in a final state yet. 
This allows us to capture different transmission strategies, such as direct push. 

We say that the run of the protocol \protocol is \emph{completed} at point in time \curtime if no further changes happen, that is $\nodestate{\node,\curtime-1}=\nodestate{\node,\curtime}$ for every node \node.
We denote the first point in time at which the protocol is completed by \completedtime. The \nameprotocol is \emph{successful} if, upon completion, all benign nodes are informed, that is $\nodesetind{\completedtime,\stateFinal}=  \nodesetind{\completedtime}$.

Intuitively, a protocol is more effective the more nodes are informed in a shorter time. 
To quantitatively evaluate this, we capture the fraction of benign nodes being informed after time \curtime, $\robust(\curtime)$, and define it as:   
\begin{equation}
    \robust(\curtime)=\frac{ \setsize{\nodesetind{\curtime,\stateFinal}}}{\setsize{\nodesetind{\curtime}}}. \label{eq:robust}
\end{equation}

\section{Security Model \& Analysis}\label{sec:security-model}
In this section, we lay down the foundation for evaluating the security of \nameprotocols by specifying the attacker model (\Cref{sec:attacker-model}) and the notion of security (\Cref{sec:security-definition}). 
In particular, we define the notion of \emph{robustness} as a means of evaluating the effectiveness and impact of various adversarial strategies.

\subsection{Attacker Model}\label{sec:attacker-model}
We start by modeling possible adversarial (i) actions and (ii) knowledge.

\subheading{Adversarial Actions.} 
We adopt the common model used in distributed systems~\cite{DBLP:conf/fc/EyalS14,DBLP:conf/ccs/GervaisKWGRC16,Dembo2020}, in which the attacker can control a fraction  \corruptfrac of the nodes to affect the protocol's execution. That is, for any point in time \curtime, the set of nodes $\nodesetind{\curtime}$ is the disjoint union of  the corrupted nodes  $\corruptednodesetind{\curtime}$ and honest nodes $\honestnodesetind{\curtime}$. 

Node corruption is modeled through the attacker’s invocation of the method \corrupt. Specifically, given a graph $\graphind{\curtime} = (\nodesetind{\curtime}, \edgesetind{\curtime})$, calling $\corrupt(\nodeset^*)$ for some node set $\nodeset^*$ results in all nodes in $\corruptednodesetind{\curtime}:=\nodesetind{\curtime} \cap \nodeset^*$ being corrupted and placed under the attacker’s control, while all remaining nodes remain honest. Notice that corrupting nodes can be subject to some restrictions:
\begin{description}
    \item[Bound: ] The fraction of corrupted nodes is bounded by parameter $\corruptfrac\in[0,1]$ with $\max_{\curtime}\left\{\frac{\abs{\corruptednodesetind{\curtime}}}{\abs{\nodesetind{\curtime}}}\right\} \leq \corruptfrac$. An attacker is naturally most effective when it corrupts as many nodes as its budget allows. Unless stated otherwise, we therefore assume that the attacker exhausts its budget and corrupts a fraction $\corruptfrac$ of nodes. 
    \item[Static vs. mobile:] A static (or monotonic) adversary fixes its set of corrupted nodes at the outset, whereas a mobile adversary may adaptively change this set at any time, subject to all other constraints. As churn reshapes the topology, nodes corrupted early lose relevance, so static corruption converges to random corruption. Unless stated otherwise, all references to adaptive adversaries therefore denote the more realistic mobile adversary. 
    \item[Delay: ]Corrupting nodes may not happen instantaneously but can be delayed by some value  \corruptdelay. That is, if an attacker issues $\corrupt(\nodeset^*)$ at epoch \curtime, the set of corrupted nodes remains unchanged until epoch $\curtime+\corruptdelay$ so that $\corruptednodesetind{\curtime+\corruptdelay} =\nodesetind{\curtime+\corruptdelay}\cap \nodeset^*.$ Note that $\corruptednodesetind{\curtime+\corruptdelay}\subsetneq \nodeset^*$ can happen if some of the nodes in $\nodeset^*$ have been removed due to churn.
\end{description}

Corrupted nodes are fully controlled by the attacker and may deviate arbitrarily from the prescribed \nameprotocol, e.g., by selectively withholding messages from peers or leaving the network altogether. Conforming to the BFT model, our model also accounts for nodes that unexpectedly crash during consensus realization. We say that a node is \emph{crashed} if it remains offline throughout an entire consensus round. The prevalence of such failures is governed by a churn rate $c$ and a distribution over nodes' offline and online durations, both of which we instantiate in Section~\ref{sec:evaluation} using empirically derived values from real-world deployments. For instance, $\corruptfrac=0$ and $\churnrate>0$ capture a graph free of active corruption yet still subject to churn and crashes. 

In this work, we focus on two specific adversarial behaviors upon receiving a rumor message:
\begin{description}
    \item[Eclipsing:] The node ignores all received messages and does not forward them. This effectively eclipses its neighbors from ongoing message propagation.
    \item[Passive non-responsive: ]The node advertises receipt of a message but withholds its content~\cite{tempering-delivery}, increasing  latency. This is especially effective in protocols that use timeouts to save bandwidth, as a well-connected corrupted node that is among the first to advertise can significantly delay delivery to its neighbors.
\end{description}

\subheading{Adversarial Knowledge.}\label{sec:adversarial-knowledge}
We assume an attacker who knows the deployed \nameprotocol, the churn parameters, and the full node set \nodeset, and who may additionally hold (partial) knowledge of the graph topology, as described below. 
Our analysis deliberately targets the robustness of the gossip layer itself, i.e., its behavior once peers have been discovered and the overlay topology has been established. Namely, peer discovery determines how a topology comes about. Gossip robustness determines what that topology can withstand. We treat these as separate concerns for three reasons. First, the two layers compose rather than interact: whatever discovery mechanism is in use, its output is a topology, which our analysis takes as input. Second, this makes our results agnostic to the peer discovery protocol in use: they hold for any protocol that yields a given topology, rather than being tied to a particular design. Third, attacks on discovery itself, such as eclipse attempts that bias neighbor selection, are therefore not excluded but subsumed: their effect is captured by the adversarial topology they produce, which our model can evaluate directly.

\emph{Minimum knowledge} refers to the attacker knowing only the connections of the corrupted nodes, while \emph{full knowledge} means that the attacker knows the topology of the full graph. 
The third, intermediate category is referred to as \emph{partial knowledge}. This setting is more realistic than the other two: we assume that the adversary can directly communicate with a subset of nodes---for example, those that can be inferred through active probing, such as address advertisement or transaction isolation~\cite{li2021toposhot,zhao2024dethna,delgado2019txprobe,cao2020MoneroTopo}---but does not necessarily know their peers or their connections. 
In \cref{sec:methodology}, we justify the assumed attacker knowledge for each of the concrete blockchain systems we consider.


\subsection{Robustness}\label{sec:security-definition}

In \cref{sec:protocol}, we introduced $\robust(\curtime)$ as a quantitative measure of a gossip protocol's effectiveness. Intuitively, robustness captures the protocol's ability to maintain $\robust(\curtime)$ under attacks.
However, two issues arise. First, these values fluctuate due to protocol randomness and effects such as churn; hence, $\robust(\curtime)$ refers to expected values. Second, comparing all individual values is often uninformative, as transient performance is less relevant. Instead, we focus on key metrics, motivated by the principle that a gossip protocol should inform \emph{as many} nodes as possible in the \emph{shortest} time, captured by the following two notions:
\begin{description}
    \item[Coverage: ] $\maxcoverage=\max_{\curtime\geq \starttime}\{\robust(\curtime)\}$ denotes the highest possible fraction of benign nodes that eventually will be informed. Ideally, \maxcoverage should be equal to or at least close to $1$. \footnote{Note that while a static adversary limits coverage to at most $\maxcoverage \leq 1-\bound$, the coverage can reach up to $\maxcoverage\geq 1-\bound$ against an adaptive adversary that can reassign its corrupted nodes during propagation.} 
    \item[Delay: ]$\maxdelay=\min\{\delta\geq 0\vert \robust(\starttime+\delta)=\maxcoverage\}$ is the minimum time it takes until the maximum coverage is achieved. 
\end{description}
Note that these two metrics are both security-critical. Lower coverage means some nodes remain uninformed, potentially breaking consensus (see also the examples given in \cref{sec:intro}). Similarly, increased delay is known to undermine security, e.g., see \cite{Dembo2020,albrecht2024larger}. An attacker targeting a gossip protocol would therefore aim to disrupt its operation by reducing \maxcoverage and/or increasing \maxdelay.
In other words, we say that an adversary \attacker successfully disrupted the coverage if $\maxcoverage'<\maxcoverage$. On the other hand, the protocol provides robust coverage if $\maxcoverage-\maxcoverage'$ is not too large.  
Similarly, an attack against the delay is successful if $\maxdelay'>\maxdelay$, and the protocol provides robust delay if $\maxdelay'-\maxdelay$ is small.

\subsection{Analysis: Random Graphs with BFT}\label{sec:Breaking-BFT-Bound}

\begin{figure*}[htb]
    \centering
    \begin{subfigure}[c]{0.32\textwidth}
        \includegraphics[width=\textwidth]{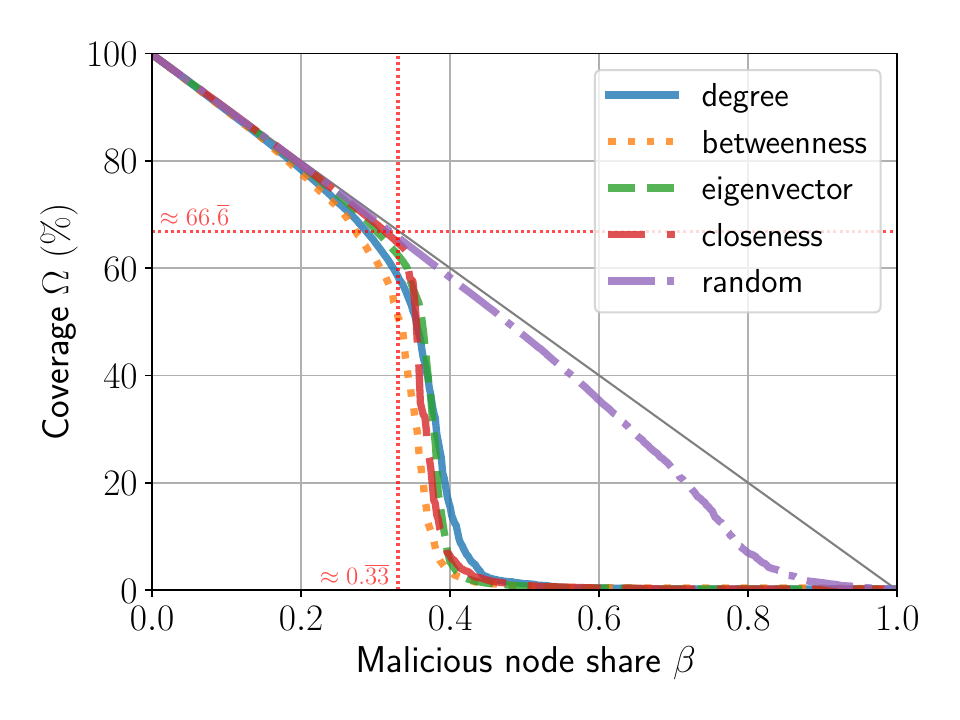}
        \caption{Varying centrality metrics, $\networksize=500$ and $\nodedegree = ln(\networksize) = ln(500)$}
        \label{fig:lcc_attack_ablation}
    \end{subfigure}
    \begin{subfigure}[c]{0.32\textwidth}
        \includegraphics[width=\linewidth]{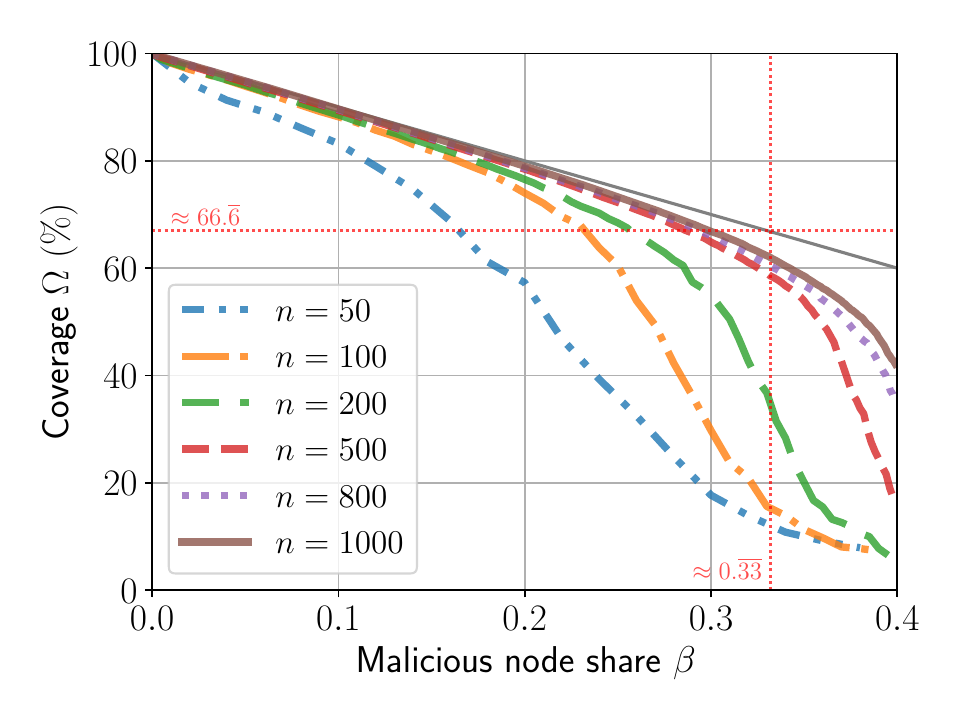}
        \caption{Varying \networksize, $\nodedegree \approx\ln(\networksize)$ ($\ERconnectivity=\frac{ln(\networksize)}{\networksize}$)}
        \label{fig:lcc_dynamic_p}
    \end{subfigure}
    \begin{subfigure}[c]{0.32\textwidth}
        \includegraphics[width=\textwidth]{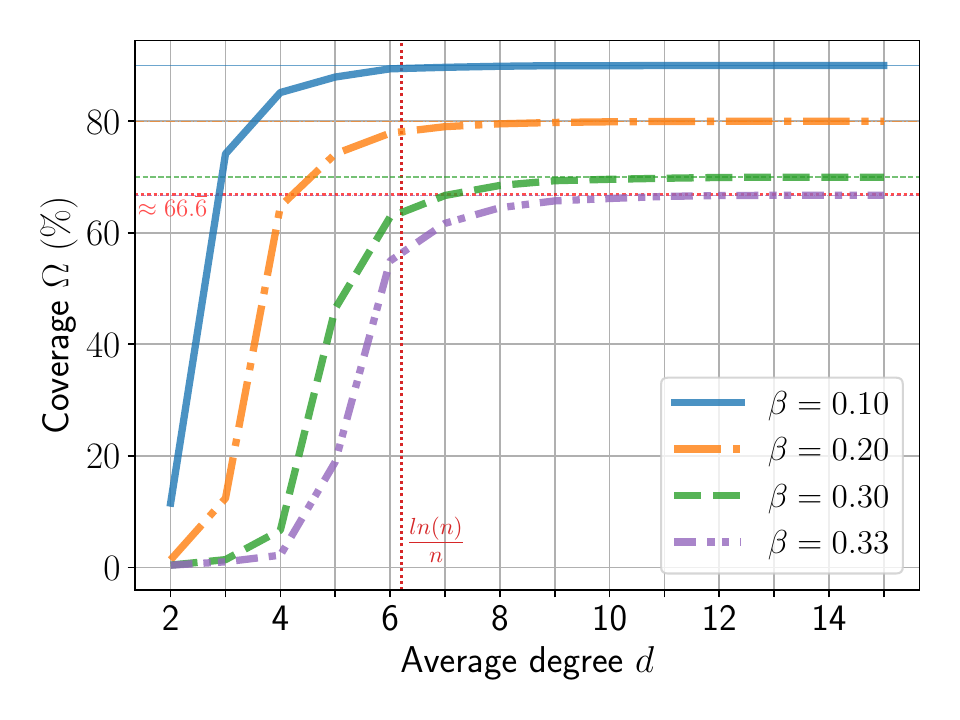}
        \caption{Fixed $\networksize=500$ and varying $\nodedegree$}
        \label{fig:lcc_degree_ablation}
    \end{subfigure}
    \caption{Coverage \maxcoverage under node corruption in Erd\H os–R\'enyi graphs without churn. The black diagonal in Figures~\ref{fig:lcc_attack_ablation} and~\ref{fig:lcc_dynamic_p} denotes the upper bound of the coverage given the malicious node share, i.e., $1-\bound$. }
    \label{fig:lcc_sizes}
\end{figure*}

We now leverage our model to analyze strategies for preventing honest nodes from receiving messages, thereby circumventing quorum formation. 
We consider a system built from a straightforward combination of gossip-based dissemination and classic BFT consensus protocols\footnote{An analogous analysis can be derived for Nakamoto-style protocols with minimum \nsboundtext = $\frac{1}{2}(\networksize+1)$ instead.}.

\begin{figure}
    \centering
    \includegraphics[width=0.87\linewidth]{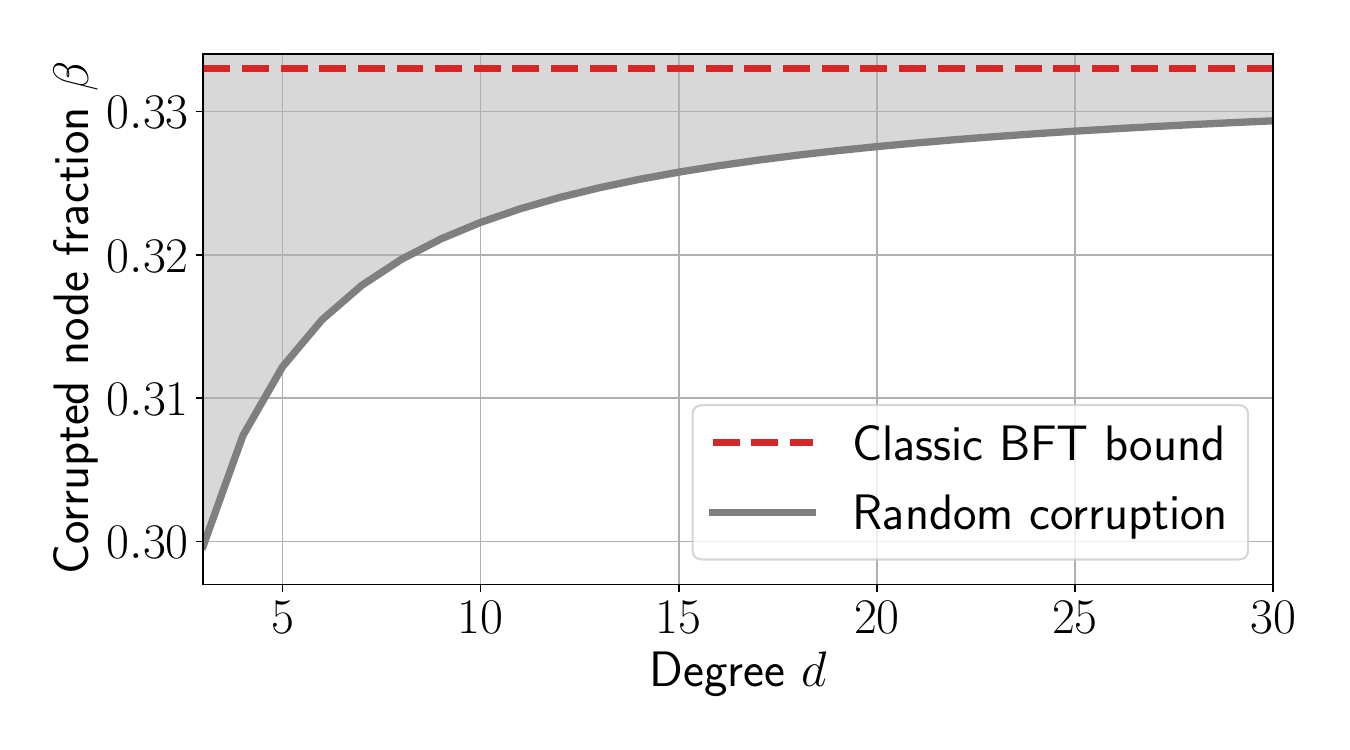}
    \caption{
        $\corruptfrac$ in the peer-corruption strategy (cf.~\cref{eq:bound_attack_global_knowledge}) as a function of the (minimum) node degree $\nodedegree \in [3,30]$ in a network of size $\networksize=\num{1000}$.  
    }
    \label{fig:eclipse_attack_corruption_bound}
\end{figure}

\subheading{Random Corruption.} 
A straightforward strategy consists of randomly corrupting \corrsize nodes, with the additional effect of isolating further nodes, beyond those directly compromised, in an untargeted manner. By isolated, we refer to a non-corrupted node that 
shares no connections with other honest nodes~\cite{eclipse-bitcoin,tempering-delivery}. Recall that, in classic BFT protocols, forming a  quorum is not possible when:

\begin{eqnarray}
\networksize-\left(\corrsize+\isosize+\crashsize\right)&<&\frac{2\networksize+1}{3}  \nonumber \\  
\Leftrightarrow\     \frac{1}{3}(1-\frac{1}{\networksize}) &<& \corruptfrac  +  \frac{\isosize}{\networksize}+  \frac{\crashsize}{\networksize},\label{eq:isoCondition}
\end{eqnarray}

\noindent where \corrsize denotes the number of actively corrupted (non-crashed) nodes, \crashsize the number of crashed nodes, \isosize the number of nodes additionally isolated by the adversary (but that did not otherwise naturally crash), and \corruptfrac the fraction of corrupted nodes. 
Eq.~(\ref{eq:isoCondition}) reduces to the classic BFT condition when $\isosize=0$.

Assuming a uniform $\nodedegree = \left|\peers{\node}\right|,\ \forall \node\in\nodeset$, the probability for any non-corrupted node \node of being isolated equals the probability that all of its $\nodedegree$ peers are corrupted, which is approximately $\corruptfrac^d$.
Hence, the expected  fraction of isolated nodes is $\frac{\isosize}{\networksize}=\corruptfrac^{\nodedegree}\cdot \left( 1-\corruptfrac\right)$.

Plugging this into Eq.~(\ref{eq:isoCondition}) shows that the expected coverage would not be enough for quorum formation when:

\begin{equation}
    \frac{1}{3} \left(1-\frac{1}{\networksize}\right)<  \corruptfrac + \corruptfrac^{\nodedegree}\cdot (1-\corruptfrac)+  \frac{\crashsize}{\networksize}.\label{eq:full-condition-random-attack}
\end{equation}
We can also see that an increase of $\nodedegree$ (towards a fully connected graph) reduces the risk of nodes getting isolated.

\subheading{Greedy Strategies.} We note that adversaries with partial or full knowledge of the graph topology can mount even more effective strategies to corrupt nodes for preventing messages from being received by a larger set of honest nodes, as illustrated in the example in \cref{fig:separable_graph_2}.

For instance, one could randomly select nodes to be isolated and then compromise all their peers. In this case, it holds that $\isosize\geq \frac{\corrsize}{\nodedegree}$ (peer sets of distinct nodes may overlap). Even if we assume to the disadvantage of the attacker that $\crashsize=0$,  Eq.~(\ref{eq:isoCondition}) provides the following sufficient criterion that this strategy, dubbed peer-corruption-strategy, prevents quorum formation in classical BFT protocols:
\begin{equation}
    \corruptfrac>\frac{\nodedegree}{3\nodedegree+1}\cdot \left( 1 - \frac{1}{\networksize}\right). \label{eq:bound_attack_global_knowledge} 
\end{equation} 
Since $\frac{\nodedegree}{3\nodedegree+1}<\frac{1}{3}$ for any $\nodedegree\geq 1$, we can see that corrupting less than $1/3\cdot (1-1/\networksize)$ of the nodes can already be sufficient to prevent quorum formation in this context (see Figure~\ref{fig:eclipse_attack_corruption_bound}). 

More generally, a strategy could be to compromise nodes based on their placement/centrality within the network, for instance, according to their degree, eigenvector, betweenness, or closeness centrality (see~\cite{Koschützki2005} and \cref{sec:centrality_metrics} for details). Importantly, estimating node centrality does not require complete topological knowledge. Public nodes, for example, naturally serve as entry points to the network and are therefore expected to have higher centrality than non-public nodes. We demonstrate how feasible such topology inference is in practice for Ethereum, Solana, and Polkadot in~\Cref{sec:methodology}.

In Figure~\ref{fig:lcc_sizes}, we analyze the coverage \maxcoverage under the greedy strategy when an adversary iteratively corrupts a fraction $\bound$ of nodes as follows: at each step, the adversary removes the currently most central node under the selected metric, updates the graph accordingly, and recomputes centrality scores on the remaining network---matching the procedure used in Section~\ref{sec:evaluation}. 
To this end, we generate 20 Erd\H os–R\'enyi (ER) graphs with sizes $\networksize \in [50,1000]$ using the NetworkX~\cite{networkx_github} Python library and numerically compute the average \maxcoverage for each parameter configuration.  Unless otherwise specified, we consider a dynamic connectivity parameter\footnote{Erd\H os and R\'enyi~\cite{Erdos_Renyi} have shown that $\ERconnectivity = \frac{\ln(\networksize)}{\networksize}$ is a threshold for an ER graph to be almost surely connected.} $\ERconnectivity = \frac{\ln(\networksize)}{\networksize}$ ($\nodedegree=\ln(\networksize)-\frac{ln(\networksize)}{\networksize} \approx \ln(\networksize)$).
As our primary interest here is the relationship between the number of corrupted nodes \corrsize and the number of isolated nodes $\isosize$, we assume $\crashsize=0$ in our analysis in Figure~\ref{fig:lcc_sizes}. Section~\ref{sec:evaluation} extends these results to more realistic settings, using empirically grounded churn rates under which a subset of nodes may appear offline or crashed.

Figure~\ref{fig:lcc_attack_ablation} illustrates the impact of the chosen centrality metric for a fixed network size of $\networksize=500$. 
Note that given $\networksize\bound $ corrupted nodes, the ideal coverage is $(1-\bound)$ (represented in Figure~\ref{fig:lcc_attack_ablation} by the straight, solid line), achieved when the honest subgraph remains fully connected. 

We clearly see that greedy strategies that leverage knowledge of node placement consistently achieve lower coverage than the random corruption strategy. Among the evaluated metrics, betweenness centrality yields slightly lower coverage than degree centrality or closeness centrality, but its computation is expensive and requires detailed knowledge of pairwise network latencies. We therefore base most of our analysis (cf.~\cref{sec:rnd_graph_robustness}) on degree centrality, which relies only on local degree information that is more realistically available to an adversary. 

Figure~\ref{fig:lcc_dynamic_p} quantifies the average \maxcoverage as a function of the fraction \corruptfrac for varying network sizes, using the same connectivity as in Figure~\ref{fig:lcc_attack_ablation} and assuming a degree centrality metric.  
Given a fixed connectivity parameter $\ERconnectivity=ln(\networksize)/\networksize$, smaller network sizes in ER graphs are associated with weaker subgraph connectivity, e.g., $\nodedegree \approx\ln(\networksize) = 3.9$ for $\networksize=50$ compared to $\nodedegree \approx 6.9$ for $\networksize=1000$. This results in a faster decline of $\maxcoverage$ below the classical BFT quorum threshold (cf. Figure~\ref{fig:lcc_dynamic_p}). 
Finally, Figure~\ref{fig:lcc_degree_ablation} evaluates \maxcoverage for networks with a fixed size of $\networksize=500$ and varying average degrees $\nodedegree \in [2,15]$. To achieve a target average degree $\nodedegree$, we set the edge probability to $\ERconnectivity = \frac{\nodedegree}{\networksize-1}$. Our results indicate that sufficiently dense networks---specifically those with an average degree of $\nodedegree \geq 12$---are highly resilient in ER graphs with $\networksize=500$ nodes.
On the other hand, in an ER graph with average degree $\nodedegree=6$, an adversary that corrupts $\bound = 0.3$ of nodes can reduce the coverage to   $62.8\%$, effectively preventing quorum formation in this context. 

\section{Robustness Evaluation}\label{sec:evaluation}

In this section, we extend our analysis and empirically evaluate coverage~$\maxcoverage$ and delay~$\maxdelay$ in real-world deployments across different \nameprotocols, including random graphs, Polkadot’s grid topology, Ethereum’s Gossipsub protocol, and Solana’s dynamic trees protocol, with varying capabilities as defined in Section~\ref{sec:model} and real-world deployment artifacts, such as churn.

\subsection{Methodology \& Setup} \label{sec:methodology}

We conduct our empirical measurements using SimBlock~\cite{simblock,albrecht2024larger}, an event-based blockchain network simulator implemented in Java. SimBlock models block propagation with configurable network topologies, inter-node latencies, and dissemination protocols.

Building on this open-source framework, we implement further extensions to model four case studies inspired by \nameprotocols used in Nakamoto-style blockchains (Bitcoin and Altcoins), Polkadot, Ethereum~2.0, and Solana: random graphs, grid topologies, GossipSub, and dynamic trees. These extensions add richer transmission models, diverse topologies, and more powerful adversaries employing multiple corruption strategies.

Our simulator takes as input the network topology that emerges once peer discovery completes and evaluates its resilience to attack. Because the simulator depends only on this final topology, our results extend to any peer discovery mechanism that produces it. Moreover, since our focus is solely on message dissemination, we abstract away block mining and instead generate messages at fixed intervals from randomly selected honest nodes.

Note that evaluating our attacks on live blockchain networks is ruled out by the ethical implications of affecting real users. Moreover, such an evaluation could not provide the evidence behind Figures~\ref{fig:er-strategy-comp}--\ref{fig:solana}, for two reasons. First, we do not control the messages sent in live networks, so comprehensive end-to-end delay measurements are infeasible. Second, parameters such as network size and churn rate cannot be varied in a controlled way on live blockchains, which makes it impossible to evaluate robustness across them.

\subheading{Simulator Validation}
We build on SimBlock~\cite{simblock}, a widely adopted blockchain simulator that is calibrated with real-world parameters and has been validated against real-world deployments~\cite{simblock,albrecht2024larger}. Our implementations faithfully follow the specifications of the respective systems, including message sizes and protocol parameters. This grounds our results in realistic conditions, in line with prior simulation-based studies~\cite{tempering-delivery, albrecht2024larger, eclipse-ethereum, DBLP:conf/fc/EyalS14, DBLP:conf/ccs/GervaisKWGRC16}.

We validate our simulator extensions in two ways. First, we show that our empirical results agree with the analytical results of Section~\ref{sec:Breaking-BFT-Bound}. Second, we compare the simulator against the real networks themselves: for each considered blockchain, we check the simulated message-arrival latency against the latency distribution observed by unmodified nodes running on the live network. Our results (see Appendix~\ref{sec:validation}) confirm that the simulator faithfully reproduces the real-world behavior of the considered blockchains.

\subheading{Communication \& Messages.}  Communication is simulated via events stored in a global priority queue ordered by execution time. The simulator processes the earliest event, advances simulated time, and schedules follow-up events as needed. For example, processing a message-reception event schedules forwarding events to neighboring nodes with delays determined by the propagation model. We record the time when a message arrives on each node and use that to derive $\robust(t)$, coverage $\maxcoverage$, and delay $\maxdelay$. 

Messages are issued every 100~s for a total of \num{1000}~messages, resulting in roughly one day of simulated time.
Unless stated otherwise, each data point is averaged over 10 independent runs.

\subheading{Nodes' Locations \& Latencies.} Recall that our network setup follows the empirically measured characteristics of Bitcoin, as calibrated in prior work~\cite{tempering-delivery, albrecht2024larger, eclipse-ethereum, DBLP:conf/fc/EyalS14, DBLP:conf/ccs/GervaisKWGRC16}. Specifically, we adopt their distribution of nodes across geographic regions, their inter- and intra-region latencies, and their bandwidth values. Messages are transmitted sequentially, with transmission time equal to the message size divided by the bandwidth plus the link latency.

\subheading{Number of Nodes \& Churn.} 
Note that the number of reachable full nodes in Bitcoin is approximately 5000  ~\cite{tempering-delivery}, $512$ for Ethereum (active committee)\footnote{See the “Sync Aggregation’’ field in block \url{https://beaconcha.in/block/22431860}.}, $\num{600}$ for Polkadot\footnote{\url{https://polkadot.subscan.io/validator}}, and $\num{1200}$ for Solana~\cite{solana-nw-size}. 
Motivated by this, we vary the network size from 50 to $\num{5000}$ nodes and mark a fraction of nodes as churn nodes $\churnrate \in [5\%, 50\%]$ that periodically join and leave.\footnote{In Bitcoin, up to 97.6\% of the nodes are affected by churn~\cite{churn}.} 
These nodes repeatedly join and leave the network. Their online durations are drawn from a log-logistic distribution $(\alpha_{distr.}=11.0,\ \beta_{distr.}=0.771)$, while their offline durations are drawn from a Weibull distribution $(\lambda=0.64,\ k=0.183)$, consistent with empirical Bitcoin measurements~\cite{churn}. After each period expires, a node toggles its online-offline state.

Note that in our experiments, we ensured that quorum formation remains possible even when nodes crash (due to churn) and the attacker uses its corruption budget. Notably, even at large churn rates $\churnrate \geq 30\%$, the fraction of simultaneously online benign nodes remains sufficiently high to form a BFT/NS quorum. For instance, in the benign setting ($\bound=0$), a churn rate of $\churnrate = 50\%$ leaves an average of approximately $83.4\%$ of nodes that are both benign and online. This is because our simulator reflects real-world churn patterns,  where online periods are significantly longer than offline periods, with medians of approximately $20.41$ hours and $7.37$ hours, respectively (a ratio of $2.77\times$). Given the same churn rate, even when \bound increases to $\bound=0.2$, we still have a majority of $66.8\%$ of all nodes being both benign and online, sufficient to form a functional quorum in classical BFT protocols.

\begin{figure}[tb]
    \centering
    \begin{subfigure}[l]{0.4\linewidth}
        \raggedright
        \hspace*{-1.8em}
        \scalebox{0.64}{
        \begin{tabular}{|c|c|} 
            \hline
            \multicolumn{2}{|c|}{Network Settings} \\
            \hline\hline
            Message size &   80KB (2KB) \\ 
            \hline  
            Average degree $\nodedegree$ &  8,15,20\\ 
            \hline 
            Topology & \makecell{Random graph, \\ Grid, \\ Dynamic tree}  \\ 
            \hline  
            Network size $\networksize$ &  \makecell{50 to 5000}\\
            \hline
            Churn rate \churnrate & 5\% - 50\% churn nodes\\
            \hline
            Message rate & 0.01  (1, 10, 120) msg/s\\ 
            \hline 
            \multicolumn{2}{c}{}\\
            \multicolumn{2}{c}{}\\
            \multicolumn{2}{c}{}\\
            \multicolumn{2}{c}{}\\
        \end{tabular}
        }
    \end{subfigure}
    \hspace{0.7em}
    \begin{subfigure}[r]{0.4\linewidth}
        \scalebox{0.64}{
        \begin{tabular}{|c|c|} 
            \hline
            \multicolumn{2}{|c|}{Adversary Settings}\\
            \hline
            \hline
            \makecell{Node corruption\\ budget $\bound$}    &  \makecell{0.1, 0.2, 0.3, \\ 0.32, 0.33, 0.5} \\ 
            \hline
            \makecell{Adversary\\adaptiveness}    &  \makecell{Static, \\ Adaptive} \\
            \hline
            \makecell{Corruption\\ strategy}      &  \makecell{None, \\ Random, \\ Greedy} \\ 
            \hline
            \makecell{Greedy corruption \\centrality metrics} &  \makecell{Degree cent., \\ Closeness cent., \\ Eigenvalue cent., \\ Betweenness cent.\ }  \\
            \hline
            Corruption delay &  \makecell{10ms to 10s}\\ 
            \hline  
        \end{tabular}
        }
    \end{subfigure}
    \captionof{table}{Summary of parameters evaluated in our setup.}
    \label{tab:simulation-params}
\end{figure}

\subheading{Topologies.}
We consider undirected network graphs\footnote{Although gossip protocols distinguish inbound and outbound links, messages are forwarded across all established connections, making propagation effectively undirected.}, and evaluate three topology classes as follows:
\begin{itemize}[leftmargin=*]
    \item \textit{Random Graphs (unstructured topology):} As explained in \cref{sec:network}, these are modeled by Erd\H{o}s–R\'enyi graphs $G(\networksize,\ERconnectivity)$.
    \item \emph{Dynamic Trees (semi-structured topology):} Inspired by Solana’s Turbine~\cite{solana-turbine}, where messages are encoded into chunks for more efficient propagation (Section~\ref{sec:case-solana}), we model a dynamic tree that is reshuffled for each message chunk. Message chunks traverse different tree paths, and the full message is reconstructed once all parts arrive.
    \item \emph{Grid Matrix (fully structured topology):} We also model the grid topology used in Polkadot~\cite{polkadot-grid}, where nodes connect to all others in the same row and column. This topology exhibits uniform degree and minimal clustering, similar to random graphs.
\end{itemize}

Note that topologies handle churn differently: random graphs resample peers for offline nodes and assign fresh connections upon return; structured topologies (grids and trees) preserve their logical structure by disabling and reactivating affected links; and GossipSub bootstraps randomly then evolves via peer scoring (Section~\ref{sec:case-gossipsub}).

\subheading{Adversaries.} We implement various adversarial strategies by scheduling attack events that compromise nodes either statically or adaptively. The adversary can select targets based on assumed topology knowledge and multiple strategies, such as random selection or greedy selection using node centrality metrics.
For simplicity, in our evaluation, we assume that the complete set of nodes is known to both honest participants and the adversary.

We vary the corruption bound over $\bound \in [0.1, 0.5)$ (cf.~Table~\ref{tab:simulation-params})\footnote{To ease presentation, $\bound=0.5$ denotes the case where the adversary compromises as many nodes as the NS-bound permits.}, and consider both static adversaries, which initially fix corruption targets, and adaptive adversaries, which adjust targets over time, subject to a nonzero corruption delay.
Overall, our evaluation spans the corruption strategies introduced in Section~\ref{sec:Breaking-BFT-Bound} and the parameter configurations summarized in Table~\ref{tab:simulation-params}. 

\begin{figure}[tbp]
    \centering
    \begin{subfigure}[c]{0.43\linewidth}
        \includegraphics[width=\textwidth]{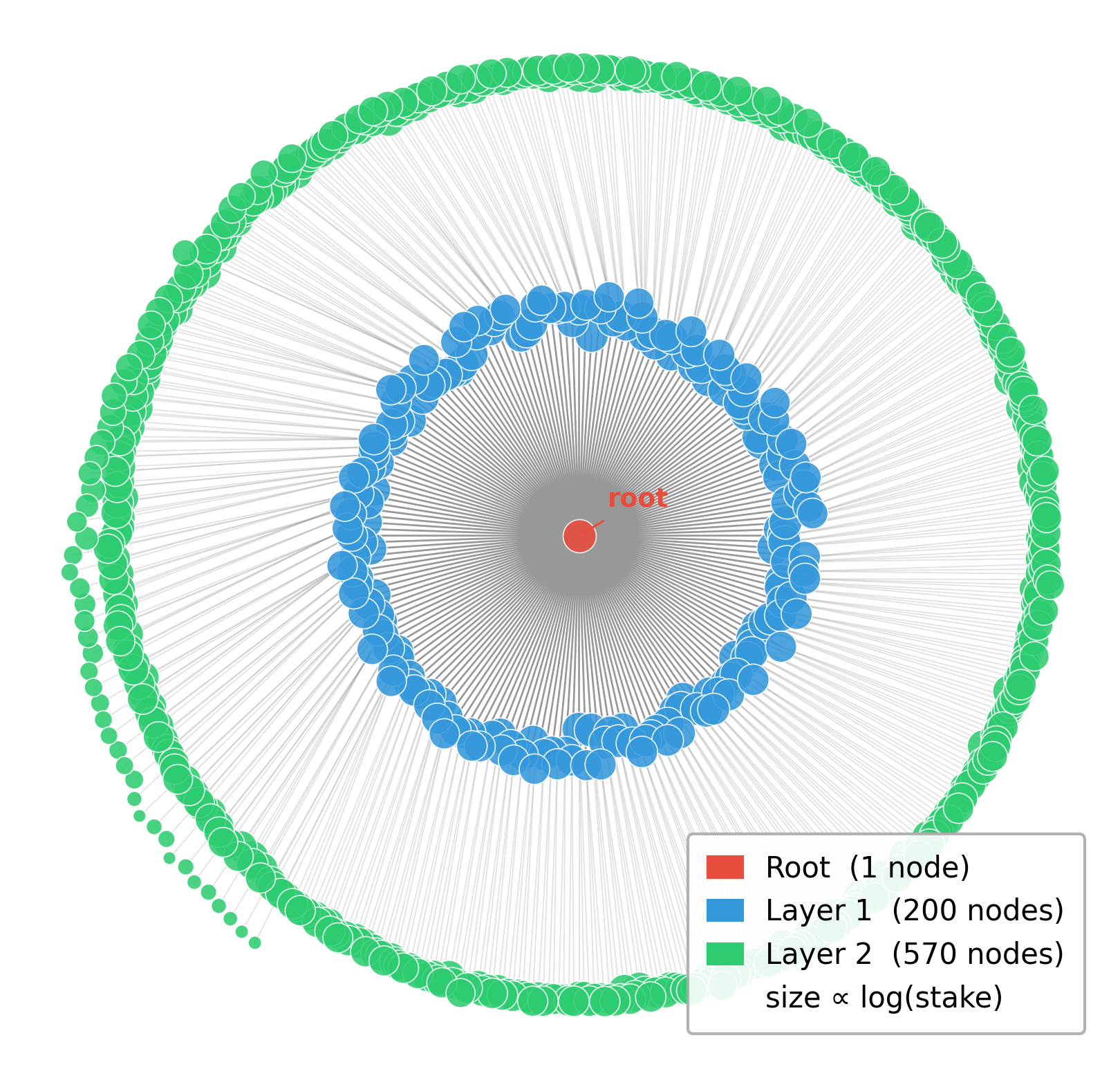}
        \caption{Snapshot of the Solana topology in slot $423812556$ and for chunk $1$. Nodes with less stake are smaller.}
        \label{fig:solana-topology}
    \end{subfigure}
    \hfill    
    \begin{subfigure}[c]{0.55\linewidth}
        \includegraphics[width=\textwidth]{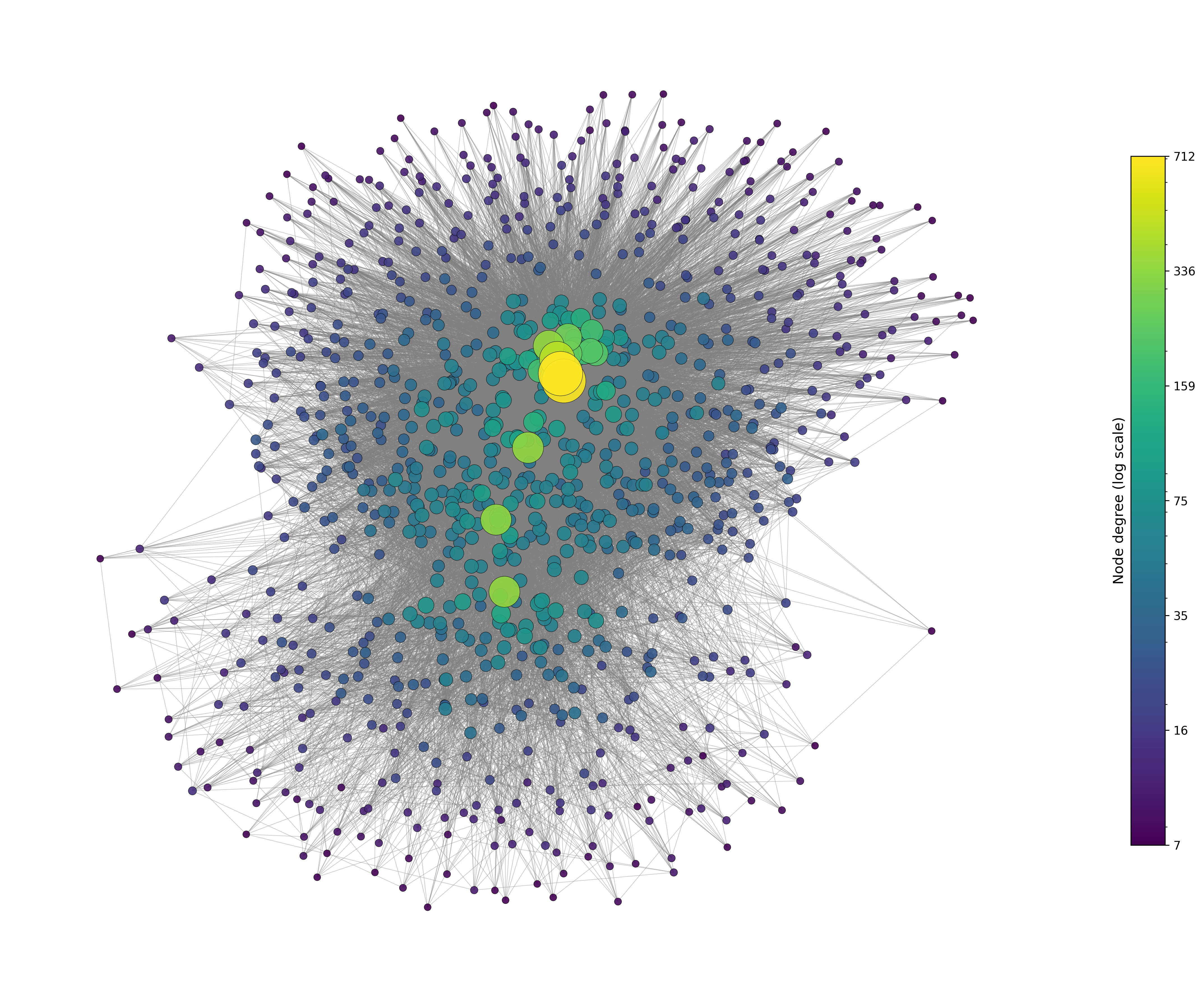}
        \caption{Topology of the largest testnet of Ethereum, Goerli, leveraging the transaction isolation approach by work~\cite{li2021toposhot}. Nodes with a higher degree are bigger and brighter.}
        \label{fig:ethereum-topology}
    \end{subfigure}
    \caption{Snapshots of the graph topologies in Solana and Ethereum testnet.}
\end{figure}

\subheading{Practical Considerations.}
The adversarial strategies that we consider assume two capabilities: (i) the ability to compromise $\corruptfrac$ of the nodes, and (ii) partial or complete knowledge of the network topology. The former is standard in consensus protocol research~\cite{Erdos_Renyi,matt2022formalizing} and rests on the attacker's ability to compromise nodes by exploiting zero-days or other vulnerabilities. 
The latter is equally justified for several reasons: First, in committee-based blockchains like Solana and Polkadot\footnote{In Polkadot, given any session index, one can retrieve the \texttt{SessionRandomness} and the corresponding \texttt{ValidatorSet} from the ledger state; the grid topology can then be reconstructed from the validator set using the \texttt{SessionRandomness} as seed and the Fisher-Yates shuffle algorithm coupled with ChaCha20 RNG.}, 
validators must know one another in order to communicate, and the protocols explicitly define these communication relationships. In both systems, validator addresses are publicly discoverable by construction. Solana's gossip layer publishes contact information tied to node identity, while Polkadot's authority discovery protocol has validators publish resolvable address records for inter-validator communication. Consequently, any entity can infer the global topology simply by accessing the ledger state. For instance, Figure~\ref{fig:solana-topology} illustrates the resulting tree structure captured on Solana in Slot $423812556$, for data chunk $1$. One can query the \code{VoteAccounts}, \code{LeaderSchedule}, and \code{ClusterNodes} ledger data, which gives information on the validator set, the leader of each slot, and the validator stake. The tree for each data chunk index $j$ in slot $i$ is fixed and predictable one epoch (about 2 days) in advance and can then be reconstructed from a shuffled validator set using the seed $(i, j, leader_i)$, the validator stakes as weights, and the ChaCha20-Poly1305~\cite{chacha20} RNG.

\begin{figure*}[tbp]
\centering
    \begin{subfigure}[c]{0.32\textwidth}
        \includegraphics[width=\textwidth]{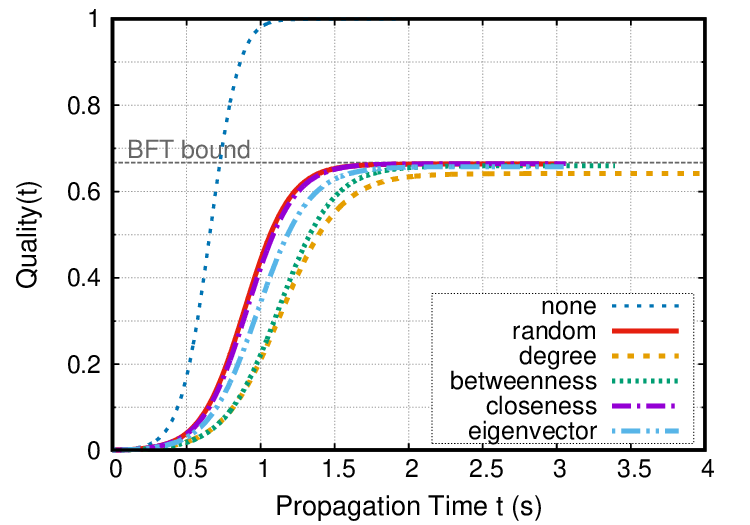}
        \captionsetup{justification=centering}
        \caption{Impact of adversarial strategy ($\networksize=1000$)}
        \label{fig:er-strategy-robust}
    \end{subfigure}
    \begin{subfigure}[c]{0.32\textwidth}
        \includegraphics[width=\textwidth]{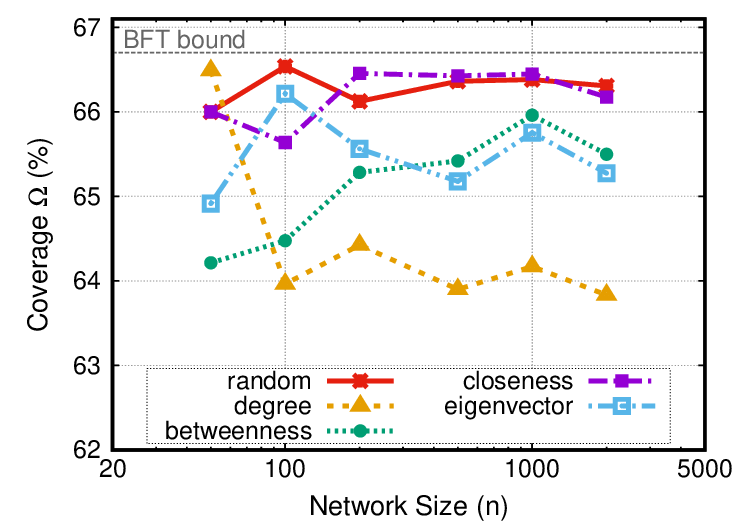}
        \captionsetup{justification=centering}
        \caption{Impact of adversarial strategy on the coverage}
        \label{fig:er-strategy-coverage}
    \end{subfigure}
    \begin{subfigure}[c]{0.32\textwidth}
        \includegraphics[width=\textwidth]{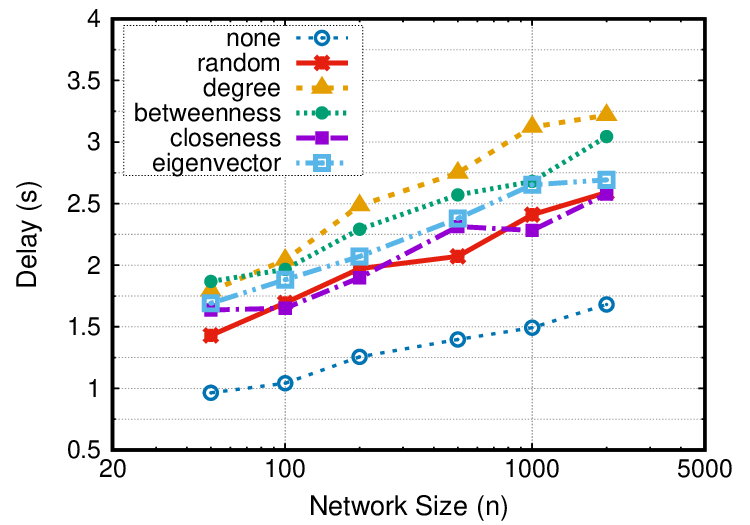}
        \captionsetup{justification=centering}
        \caption{Impact of adversarial strategy \\on the delay}
        \label{fig:er-strategy-delay}
    \end{subfigure}
    \caption{ER graphs: Impact of different adversarial strategies with $\corruptfrac=0.33$ when $\nodedegree=8$.}
    \label{fig:er-strategy-comp}
\end{figure*}

Ethereum differs from these committee-based systems in that validator identifiers are deliberately decoupled from node addresses at the protocol level. Nevertheless, discv5, the UDP-based peer-to-peer discovery protocol used by Ethereum, makes the set of beacon nodes readily enumerable. 
Alternatively, the topology can be inferred without direct participation through probing techniques such as address advertisement or transaction isolation~\cite{li2021toposhot,zhao2024dethna,delgado2019txprobe,cao2020MoneroTopo,DBLP:conf/uss/HeimbachVVKW25}. Recent measurement studies demonstrate that an unprivileged adversary can recover the remaining validator-to-node mapping in practice. 
For instance, Figure~\ref{fig:ethereum-topology} depicts the topology of Goerli, Ethereum's largest testnet\footnote{For ethical reasons, we opt to only show the Ethereum Testnet topology, as reconstructed via transaction isolation by~\cite{li2021toposhot}.}. With $1355$ nodes, Goerli exhibits a pronounced skew in node degree distribution. Note that Figure~\ref{fig:ethereum-topology} retains only nodes with degree $d \ge 8$, yet many already exceed 100 connections under the default setting of $d = 8$, with some super-nodes reaching upwards of 700 connections. As we demonstrate in Section~\ref{sec:case-gossipsub}, a greedy adversary that compromises these nodes would considerably reduce coverage in Ethereum. Moreover, most validators are nowadays cloud-hosted, allowing cloud providers to easily infer the topology. 

Bitcoin differs from committee-based settings in that a portion of its relay population operates from behind NAT and is therefore not directly reachable by an adversary. This limitation does not materially affect our analysis, however, because the nodes most beneficial to target are those holding mining power, which are necessarily reachable and publicly identifiable to participate effectively in mining. Concretely, a miner that cannot be reached can neither obtain work nor propagate its blocks competitively, and mining pools openly advertise their infrastructure. Non-mining nodes, by contrast, hold no validation power and are therefore less attractive targets. Moreover, since such nodes relay blocks and transactions over their own outbound connections, they become visible to an adversary present in the network even when they cannot be contacted directly~\cite{li2021toposhot,zhao2024dethna,cao2020MoneroTopo}. 
Their residual effect is therefore confined to the propagation delay they add along paths between miners, which prior work~\cite{DBLP:conf/ccs/GervaisKWGRC16} quantifies as negligible for Bitcoin.

Finally, note that our adversarial strategy does not require validators to be publicly identifiable or deanonymized. It only requires knowledge of the topology and node IP addresses. Even when the latter are not publicly available, an adversary that compromises a validator can directly learn the IP addresses of its neighbors, thereby obtaining partial topology information and, through repeated compromises or observations, expanding its view of the network.

\subsection{Robustness of Random Graphs} \label{sec:rnd_graph_robustness}

We evaluate the parameter combinations listed in \cref{tab:simulation-params} assuming a direct push transmission model. Networks are instantiated as Erd\H{o}s–R\'enyi graphs $G(\networksize,\ERconnectivity)$ with an average degree $\nodedegree\in \{8,15,20\}$, where $\ERconnectivity=\frac{\nodedegree}{\networksize-1}$ (cf. \cref{sec:network}).

\subheading{Simulator Validation.}
We validate the simulator by comparing empirical results with the analytical bounds from Section~\ref{sec:Breaking-BFT-Bound}.
Greedy corruption strategies significantly outperform random corruption (Figures~\ref{fig:lcc_attack_ablation},~\ref{fig:er-strategy-robust}). Targeting high-degree or high-betweenness nodes is most effective, followed by eigenvector and closeness-based strategies. Degree-based corruption consistently yields lower coverage than betweenness-based attacks, while being computationally cheaper; we therefore focus on degree-based adversaries.
Higher average degree improves robustness, leading to higher coverage and shorter delays (Figure~\ref{fig:lcc_degree_ablation}, Figure~\ref{fig:er-degree-comp} in Appendix~\ref{sec:er_add_plots}). For $\networksize=\num{5000}$ and $\bound=0.5$, coverage increases from $36\%$ to $49.98\%$ and delay drops from $6.3$~s to $4.8$~s as $\nodedegree$ grows from 8 to 20. Overall, the simulator faithfully reproduces the effects described in Section~\ref{sec:Breaking-BFT-Bound}.

\subheading{Impact of Adversarial Strategy.}
We now compare random corruption (emulating minimum knowledge about the graph topology) and greedy adversarial strategies (emulating partial/full knowledge). 
We compare the impact of the adversarial corruption bound $\bound$ on the coverage required by the minimum \bftboundtext ($\frac{1}{3}(2\networksize+1)$) or the Nakamoto-style quorum size (\nsboundtext) ($\frac{1}{2}(\networksize+1)$). 
As shown in Figure~\ref{fig:er-strategy-comp}, degree- and betweenness-based attacks most strongly degrade coverage and delay. For $\networksize=\num{1000}$ and $\corruptfrac=0.33$, degree-based corruption doubles delay (from $1.49$~s to $3.12$~s) and reduces coverage to $64.17\%$.

By varying $\bound$, we observe that for a fixed node degree of $\nodedegree = 8$, coverage is largely independent of network size (Figure~\ref{fig:er-strategy-coverage}), but decreases nonlinearly as $\bound$ increases (Figure~\ref{fig:er-bound-robust} and Figure~\ref{fig:er-bound-coverage} in Appendix~\ref{sec:er_add_plots}), consistent with known partitioning behavior in ER graphs. For $\networksize = \num{1000}$ and $\nodedegree = 8$, coverage drops from 68.15\% at $\bound = 0.3$, which is sufficient for the \bftboundtext, to 65.51\% at $\bound = 0.32$, falling below the required threshold. 
At $\bound = 0.5$, coverage further declines to only 35.61\%, well below what is necessary for safe consensus.

We also note that delays grow logarithmically with network size and increase sharply with $\bound$ (Figure~\ref{fig:er-bound-delay} in Appendix~\ref{sec:er_add_plots}). 
For $\bound=0.5$, delay rises from $1.08$~s to $2.88$~s in a 50-node network and from $1.88$~s to $6.03$~s in a \num{5000}-node network, indicating that larger networks suffer disproportionately higher latency under attack.

\begin{figure*}[tbp]
\centering
    \begin{subfigure}[c]{0.32\textwidth}
        \includegraphics[width=\textwidth]{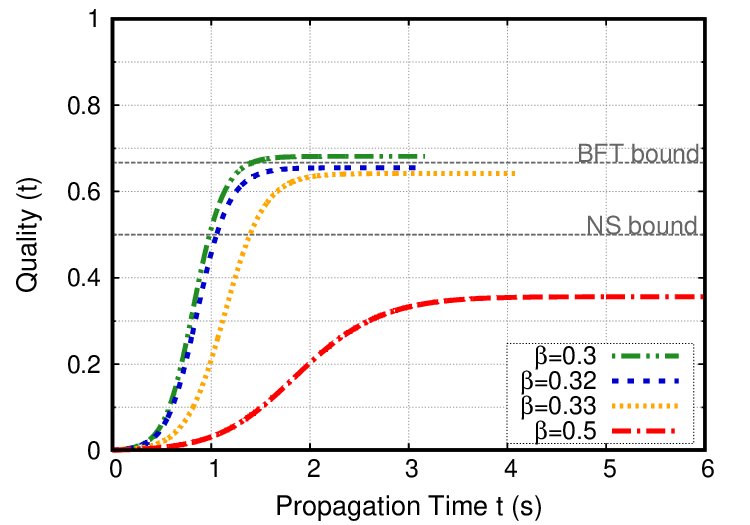}
        \caption{Impact of $\bound$ for node degree $\nodedegree=8$. }
        \label{fig:er-bound-robust}
    \end{subfigure}
    \begin{subfigure}[c]{0.32\textwidth}
        \includegraphics[width=\textwidth]{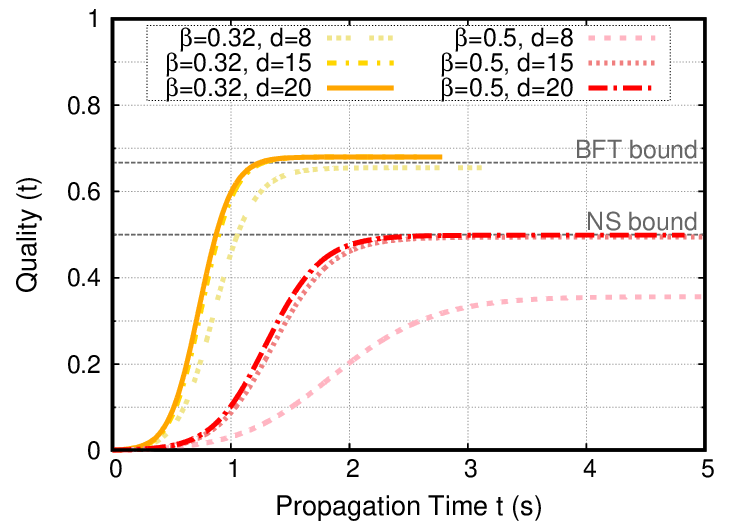}
        \caption{Impact of $\nodedegree$ and $\bound$. }
        \label{fig:er-degree-robust}
    \end{subfigure}
    \begin{subfigure}[c]{0.32\textwidth}
        \includegraphics[width=\textwidth]{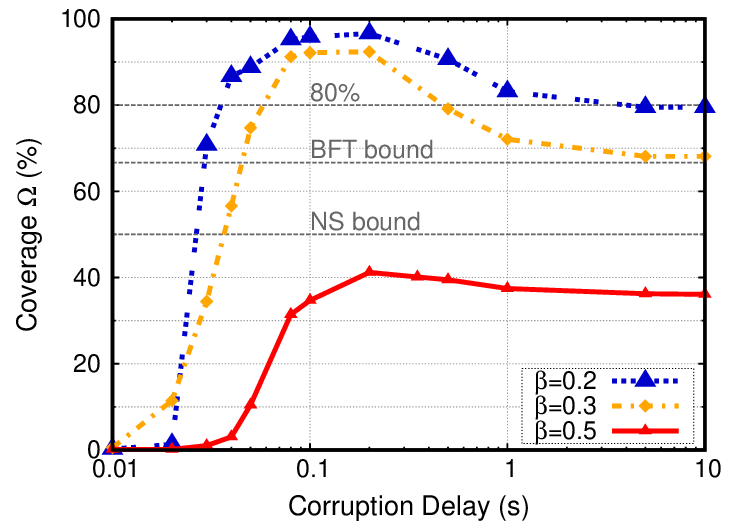}
        \caption{Impact of $\corruptdelay$ for node degree $\nodedegree=8$.}
        \label{fig:er-adapt-coverage-delay}
    \end{subfigure}
    \caption{ER graphs: Impact of $\bound$, $\nodedegree$, and $\corruptdelay$ on \namerobust for $\networksize=\num{1000}$ nodes.}
    \label{fig:er-comp}
\end{figure*}

\subheading{Impact of Node Degree.}
Higher node degrees improve robustness  (Figure~\ref{fig:er-degree-robust} and Figure~\ref{fig:er-degree-comp} in Appendix~\ref{sec:er_add_plots}). For $\bound = 0.32$, networks with $\nodedegree\in\{15,20\}$ significantly improve the robustness compared to $\nodedegree=8$.  
This trend continues under stronger attacks ($\bound=0.5$), where degree becomes decisive: in a \num{5000}-node network, coverage rises from $36.04\%$ to $49.98\%$ and delay falls from $6.27$~s to $4.8$~s as $\nodedegree$ increases from 8 to 20. 
However, even for adversaries with $\bound=0.5$, $d=15,20$ still achieve comparable performance.
Thus, beyond a threshold of $\nodedegree\approx15$, additional edges yield little robustness benefit.

\subheading{Impact of Adversarial Capability.} 
We evaluate the impact of adaptive adversaries that corrupt nodes predicted to receive messages, after a delay $\corruptdelay$ from 10~ms to 10~s
 (Figure~\ref{fig:er-adapt-coverage-delay} and Figure~\ref{fig:er-adapt-cmp} in Appendix~\ref{sec:er_add_plots}) 
and we observe that adaptive adversaries consistently increase delay and achieve varying coverage as $\corruptdelay$ grows.
With a very short corruption delay $\corruptdelay=10$~ms, adaptive attacks fully suppress propagation in all cases, confirming known impossibility results.
As $\corruptdelay$ increases, coverage improves and may exceed the maximum achievable coverage by a static adversary; e.g., with $\corruptfrac=0.3$ and $\corruptdelay=100$~ms in a \num{1000}-node network, coverage approaches $92\%$. This occurs because an adversary that repeatedly releases and corrupts different nodes over time can cause the message to traverse more nodes than under a static strategy. However, as $\corruptdelay$ increases ($\geq 5$~s), the coverage decreases and converges to the static case.
Therefore, under a smaller adversarial bound \corruptfrac, an adaptive adversary must corrupt nodes very rapidly to reduce coverage below the \bftboundtext. In our simulation settings, this requires corruption delays of approximately 20 ms for $\corruptfrac = 0.2$ and 40 ms for $\corruptfrac = 0.3$.

Our results further confirm that the mobile adversary is a more appropriate model for our analysis. As churn reshapes the topology, nodes corrupted early gradually lose their strategic value, so monotonic corruption converges toward random corruption. Figure~\ref{fig:er-adapt-coverage-delay} captures this transition by varying the corruption delay \corruptdelay: long delays approximate a monotonic adversary. Coverage plateaus once $\corruptdelay \geq 1$~s, showing that a monotonic adversary is markedly less effective than its mobile counterpart.

\subheading{Impact of Churn.}
We model network churn affecting up to $5\%$--$50\%$ of nodes (Figure~\ref{fig:er-bound-churn-coverage}) and used empirically observed patterns from~\cite{churn}  to simulate the nodes' online and offline behavior. Since message propagation completes within approximately $5$~s---significantly shorter than the typical duration for being offline---nodes that go offline miss dissemination and must rely on the consensus protocol for recovery. Even in the absence of an adversary, churn observed in 50\% of nodes reduces coverage to approximately $85\%$. 
When $\bound = 0.2$ and churn reaches $c=40\%$, coverage drops further to approximately $65\%$, falling well below the \bftboundtext. This occurs despite the fact that $75\%$ of benign nodes remain online under these conditions, a fraction that would be sufficient to form a BFT quorum if all such nodes were mutually reachable (cf.~Table~\ref{tab:avg_online_stats} in Appendix~\ref{sec:er_add_plots}).

More generally, the maximum $\bound$ that allows benign online nodes to satisfy a given quorum threshold decreases as churn increases. For example, satisfying the minimum \bftboundtext requires $\bound \le 0.2$ at $c = 30\%$ churn (with $\approx 72\%$ of nodes being benign and online), whereas satisfying the minimum \nsboundtext requires $\bound \le 0.33$ at $c = 50\%$ churn (with $\approx 58\%$ of nodes being benign and online). In these dynamic settings, average node degree has little effect on the outcome (cf.~Figure~\ref{fig:er-bound-churn-coverage}).

\begin{figure*}[tbp]
\centering
    \begin{subfigure}[c]{0.24\textwidth}
        \includegraphics[width=\textwidth]{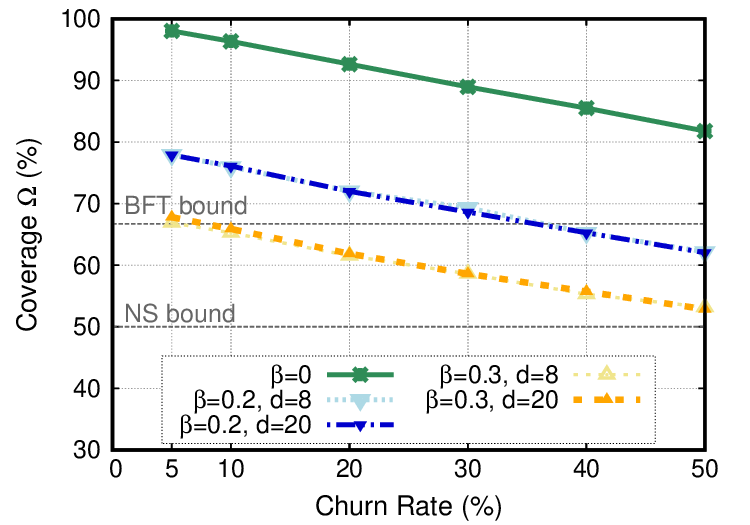}
        \captionsetup{justification=centering,margin=2em}
        \caption{ER Graphs} 
        \label{fig:er-bound-churn-coverage}
    \end{subfigure}
    \begin{subfigure}[c]{0.24\textwidth}
        \includegraphics[width=\textwidth]{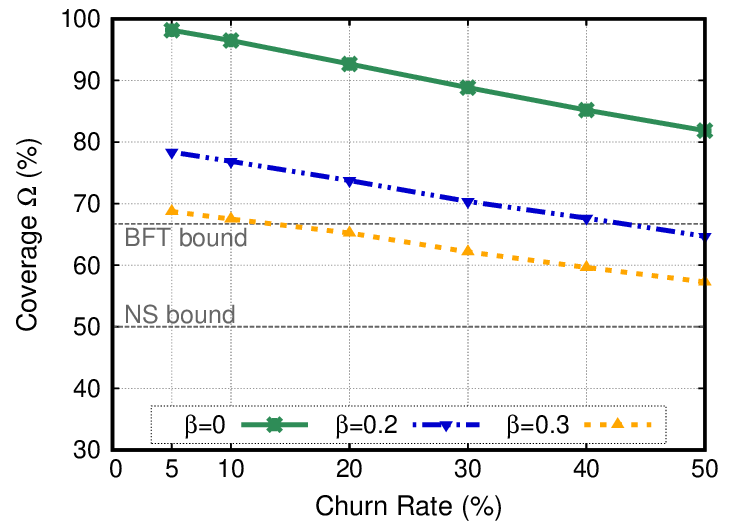}
        \captionsetup{justification=centering,margin=2em}
        \caption{Grid topologies}
        \label{fig:polkadot-bound-churn-coverage}
    \end{subfigure}
    \begin{subfigure}[c]{0.24\textwidth}
        \includegraphics[width=\textwidth]{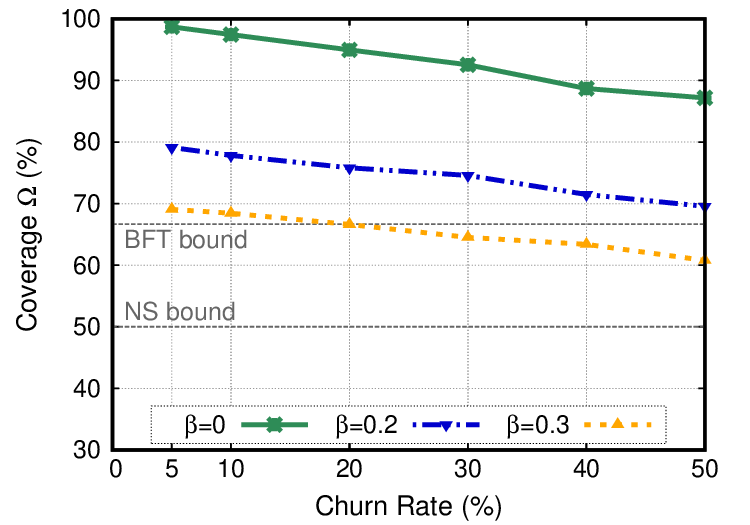}
        \captionsetup{justification=centering,margin=2em}
        \caption{GossipSub}
        \label{fig:gossipsub-churn-coverage}
    \end{subfigure}
    \begin{subfigure}[c]{0.24\textwidth}
        \includegraphics[width=\textwidth]{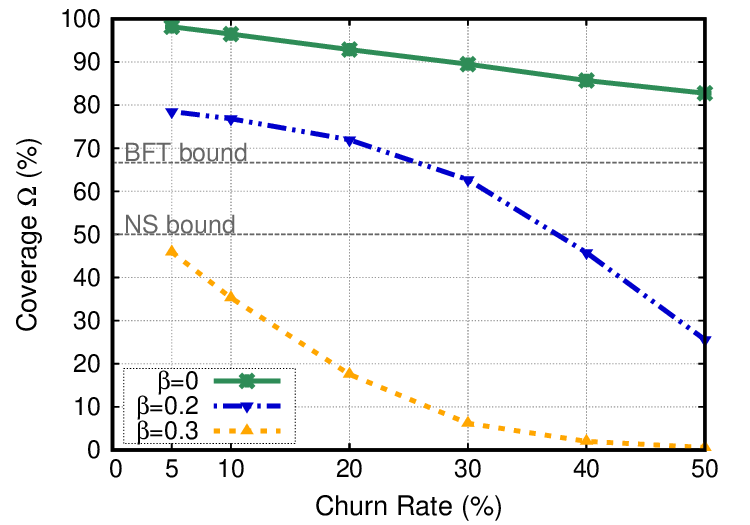}
        \captionsetup{justification=centering,margin=2em}
        \caption{Solana ($\nodedegree=200$)}
        \label{fig:solana-churn-coverage}
    \end{subfigure}
    \caption{Impact of churn rate $\churnrate$ on the coverage of different topologies with $\networksize=\num{1000}$ nodes. }
    \label{fig:diffchurn}
\end{figure*}

\subsection{Robustness of Grid Topologies  (Polkadot)} \label{sec:polkadot_results} 

Polkadot uses a nominated Proof-of-Stake protocol in which a fixed validator set is elected at the start of each 24-hour era based on stake drawn from the finalized ledger state. To disseminate messages, validators are organized into a grid topology: their IDs are randomly permuted and arranged into a $\lfloor \sqrt{\networksize} \rfloor \times \lceil \networksize / \lfloor \sqrt{\networksize} \rfloor \rceil$ matrix~\cite{grid-topo-algo}, where each validator connects to all others in its row and column, resulting in a degree of $\approx 2\sqrt{\networksize}$. Messages are forwarded to row and column neighbors, with eight randomly selected nodes additionally  
included to improve robustness. Unpredictability stems only from node churn and per-message random neighbors, since the validator set and topology are globally known and fixed within each era, and all validators have nearly identical degrees. We therefore model a static adversary with global knowledge that corrupts nodes uniformly at random. Here, messages are disseminated via direct neighbor pushes, supplemented by eight additional random peers per message.\footnote{Figure~\ref{fig:additional_nodes_ablation_Polkadot} in Appendix~\ref{sec:er_add_plots} varies the number of randomly selected nodes that each node contacts in addition to its row and column neighbors, and reports the resulting coverage and delay. The trends observed in Figure~\ref{fig:polkadot} hold across all values of this parameter, showing that it does not affect our findings.}

\subheading{Impact of Adversarial Bounds \corruptfrac.}
Figures~\ref{fig:polkadot-bound-robust}–\ref{fig:polkadot-bound-delay} show propagation coverage and delay for varying network sizes and adversarial bounds in static networks. Coverage is always at the respective bound ($1-\bound$), indicating that the remaining honest nodes stay connected despite corruption. This robustness stems from the grid’s high connectivity, with each node maintaining roughly $2\sqrt{\networksize}$ links.

The delay increases with the adversarial bound \corruptfrac, as corrupted nodes effectively remove edges and lengthen paths, mirroring trends in random graphs (Figure~\ref{fig:er-bound-delay} in Appendix~\ref{sec:er_add_plots}). For example, in a 500-node network (comparable to Polkadot’s number of validators) delay increases from about 1.55~s without attacks to 2.28~s with 
corruption $\beta=0.33$. 
We also observe that delay grows logarithmically with network size, contradicting the claim that any message traverses at most two hops~\cite{grid-topo-algo}. While any two nodes are indeed within two hops, full dissemination still requires sequential transmissions over individual connections, with the overall delay constrained by node bandwidth.

High connectivity also increases message duplication and marginally raises delays. Compared to random graphs ($\nodedegree=15$, $\bound=0.33$), grid topologies show slightly higher delays without adversaries but become more efficient under attack. For example, at $\networksize=50$, delays are 1.08~s vs.\ 0.96~s without corruption, but 1.63~s vs.\ 1.79~s at 33\% corruption; at $\networksize=\num{2000}$, the corresponding figures are 2~s vs.\ 1.68~s and 2.69~s vs.\ 2.77~s.

\subheading{Impact of Churn.}
Figure~\ref{fig:polkadot-bound-churn-coverage} shows that churn reduces coverage, consistent with results for random graphs. Due to its higher degree and additional random propagation (e.g., $\nodedegree=64$ for $\networksize=\num{1000}$ versus $\nodedegree=20$ in ER graphs), the grid achieves slightly higher coverage, especially in the presence of higher dynamics: 
from about 0.45\% improvement for $\bound=0.2$ and a churn rate of $c=5\%$, 
and up to 4.5\% for $\bound=0.3$ and $c=50\%$ churn rate.  

Overall, results from both static and dynamic networks show that the grid topology's high connectivity provides strong resilience against adversarial behavior. However, at high churn rates, it fails to guarantee the required coverage against strong adversaries ($\bound \in [0.25, 0.33]$).

\subsection{Robustness of GossipSub} \label{sec:case-gossipsub}

Ethereum 2.0, Filecoin, and Algorand all use GossipSub~\cite{gossipsub}, a publish/subscribe protocol in which nodes subscribe to \emph{topics} and receive only the messages published to them. We abstract away topics by considering a single topic to which all nodes subscribe.  
GossipSub maintains two types of connections: full-message links, over which messages are eagerly pushed to form the \emph{mesh network}, and metadata-only links, over which message identifiers are periodically advertised in batches and full messages may be lazily pulled on demand.
Each node targets a fixed mesh degree, dynamically maintained within predefined bounds: connections are added when the degree falls below the lower bound and pruned when it exceeds the upper bound. In Ethereum's default configuration, the target mesh degree is $d=8$, and batched metadata are periodically sent to 12 additional random nodes\footnote{We also vary $D_{\text{lazy}}$, the number of randomly chosen nodes in GossipSub's lazy pull mechanism (see Figure~\ref{fig:additional_nodes_ablation_GossipSub} in Appendix~\ref{sec:er_add_plots}). The results are highly consistent across all settings, showing that this parameter does not affect our findings.}, yielding a combined effective degree of $\approx 20$. Nodes also maintain per-peer scores based on live statistics, pruning peers with negative scores; further details on peer scoring are provided in Appendix~\ref{ap:peerscore}.

\begin{figure}
\centering
    \begin{subfigure}[l]{0.49\linewidth}
        \includegraphics[width=\textwidth]{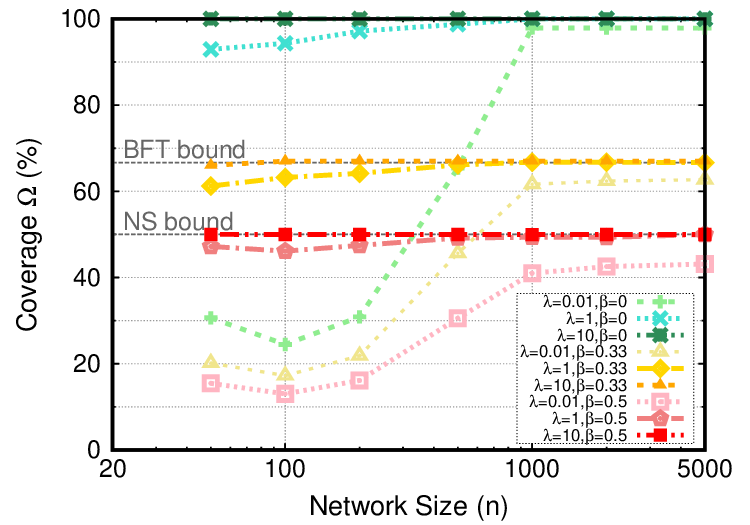}
        \caption{Impact of message rate $\arrivalrate$ on coverage (80KB)}
        \label{fig:gossipsub-bound-coverage}
    \end{subfigure}
    \begin{subfigure}[r]{0.49\linewidth}
        \includegraphics[width=\textwidth]{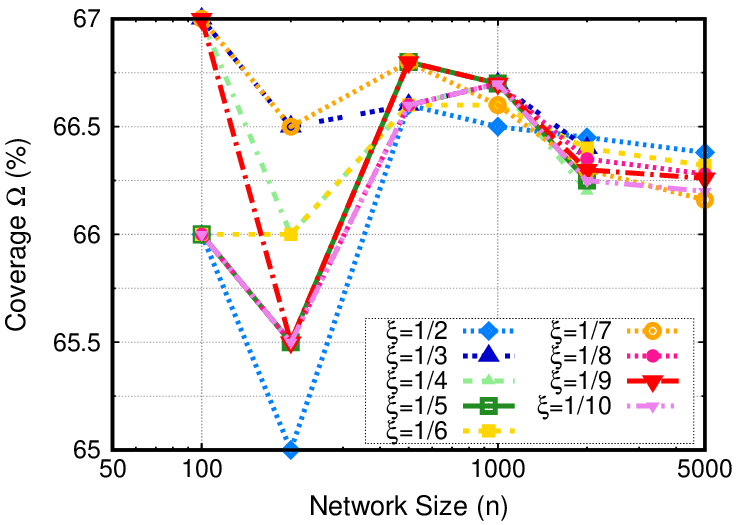}
        \caption{Impact of $\xi$ on coverage ($\arrivalrate=120$ tx/s, 2KB)}
        \label{fig:gossipsub-droprate}
    \end{subfigure}
    \caption{GossipSub: Impact of $\arrivalrate$ and $\xi$ with $\bound = 0.33$.}
    \label{fig:gossipsub-robust}
\end{figure}

Our evaluation follows the parameters used in the GossipSub evaluation report~\cite{gossipsub-eval-report} (see Table~\ref{tb:gossipsub-params} in Appendix~\ref{ap:peerscore}), which uses the libp2p Go implementation~\cite{libp2p-go-impl} 
at a rate of sending 2~KB messages of $\arrivalrate=120$ transactions per second. 
In addition, we evaluate block-sized messages of 80~KB and vary the message rate among 0.01, 1, and 10 messages per second over a \num{10000}-second simulation, corresponding to 100, \num{10000}, and \num{100000} blocks. This allows us to study how peer scoring parameters behave under different workloads.

\subheading{Impact of Message Rate.}
We first analyze a static network with large messages (80~KB) and low message rates (Figure~\ref{fig:gossipsub-bound-coverage}). For small networks ($\networksize \le 200$), coverage drops to nearly 10\% when the message rate is 0.01 msg/sec, even with no adversaries ($\bound=0$). This occurs because the peer scoring function penalizes honest nodes. In particular, one indicator (see $P_{3a}$ in Appendix~\ref{ap:peerscore}) in the peer scoring function expects each connection to relay known messages within a short window. Given the default parameters used in GossipSub (Table~\ref{tb:gossipsub-params} in Appendix~\ref{ap:peerscore}), each peer must relay at least 10 known messages per second, each within 5~ms, to avoid penalties.
At low message rates and with larger message sizes, this condition is often violated, causing honest peers to be pruned. Over time, nodes exhaust replacement candidates and eventually lose all connections, leading to poor coverage. Although coverage improves for $\networksize \ge 500$, this is only because unscored peers remain available near the end of the simulation; ultimately, connections are still exhausted. As the message rate increases, coverage improves and reaches optimal levels when $\arrivalrate \ge 10$. 

\subheading{Impact of Adversarial Strategy.}
These observations reveal that adversaries can exploit the peer scoring mechanism by selectively dropping messages rather than dropping all of them. To evaluate this strategy, we simulate the original setting in~\cite{gossipsub-eval-report} with 2~KB messages at 120 msg/s and adversarial bound $\bound=0.33$. We vary the adversarial drop rate $\droprate$ between 1/10 and 1/2 (every $10^{th}$ message and every $2^{nd}$ message is dropped) and show the \emph{minimal} achieved coverage of those dropped messages in Figure~\ref{fig:gossipsub-droprate}. 
We observe that coverage lies at the margin of the required \bftboundtext, and occasionally drops below the required threshold in certain cases. 
In particular, for small networks with $\networksize = 100$ ($\maxcoverage = 66\%$) and $\networksize = \num{200}$ ($\maxcoverage = 65\%$) at a drop rate of 0.5, a small number of honest nodes can be isolated, failing to meet the required coverage and preventing the formation of a quorum. Even in larger networks, the measured coverage often falls below the \bftboundtext.
Note that an adversary that drops all messages ($\droprate = 1$) triggers detection and subsequent pruning in GossipSub. All other evaluated drop rates, $\droprate \in [1/10, 1/2]$ incur little impact on coverage (Figure~\ref{fig:gossipsub-droprate}).

\begin{figure*}[tbp]
\centering
    \begin{subfigure}[c]{0.32\textwidth}
        \includegraphics[width=\textwidth]{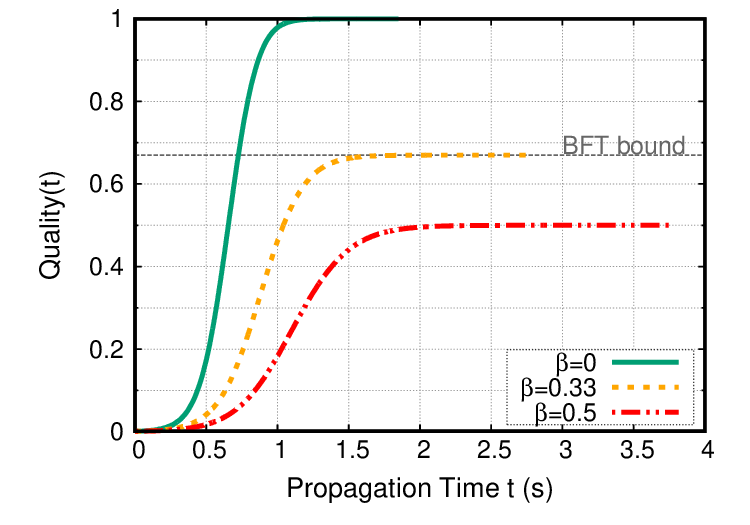}
        \caption{Impact on \namerobust ($\networksize=1000$)}
        \label{fig:polkadot-bound-robust}
    \end{subfigure}
    \begin{subfigure}[c]{0.32\textwidth}
        \includegraphics[width=\textwidth]{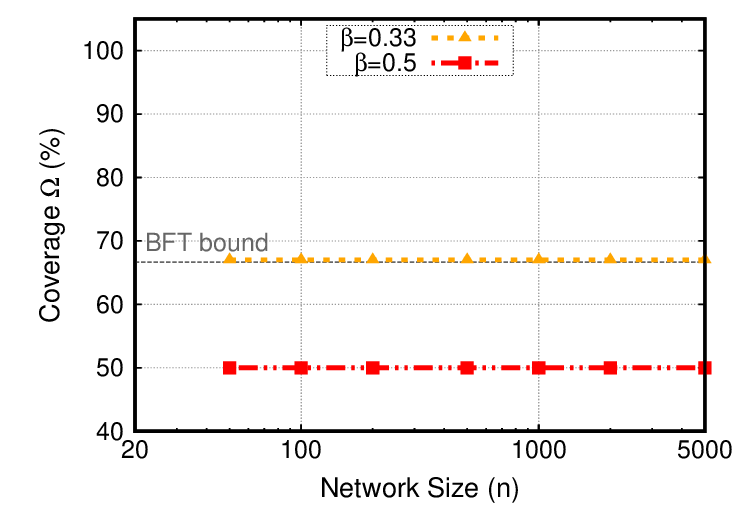}
        \caption{Impact of $\bound$ on coverage}
        \label{fig:polkadot-bound-coverage}
    \end{subfigure}
    \begin{subfigure}[c]{0.32\textwidth}
        \includegraphics[width=\textwidth]{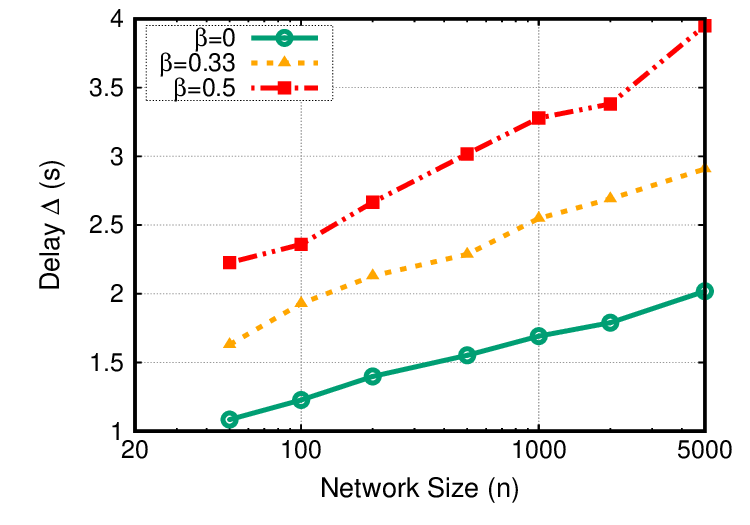}
        \caption{Impact of $\bound$ on delay}
        \label{fig:polkadot-bound-delay}
    \end{subfigure}
    \caption{Polkadot (grid topologies): Impact of $\bound$ on the robustness under random corruption.}
    \label{fig:polkadot}
\end{figure*}

\subheading{Impact of Churn.}
Figure~\ref{fig:gossipsub-churn-coverage} shows that coverage decreases as network churn increases, leading to violations of the \bftboundtext for $\bound=0.33$ at churn rates of $\geq 20\%$. Despite employing a random mesh with degree $\nodedegree=8$, GossipSub outperforms a pure random graph under identical conditions, where the \bftboundtext is violated for any churn rate $\geq 5\%$ when $\bound=0.33$. This improvement is attributable to GossipSub’s lazy pull mechanism, which periodically advertises recent message metadata to a random set of $D_{\text{lazy}}=12$ peers. As a result, coverage under churn is comparable to that of a grid topology, while avoiding excessive message overhead.

\subsection{Robustness of Dynamic Trees (Solana)} \label{sec:case-solana}
Solana’s gossip protocol, Turbine~\cite{solana-turbine}, uses a dynamic tree topology with chunk-based push dissemination. When a block is generated in an epoch, it is encoded into multiple chunks. For each chunk, a random seed derived from the slot number and chunk index constructs a new spanning tree with degree 200. The block\footnote{Solana does not define explicit blocks; instead, it produces a stream of transaction entries. We treat a batch of entries as a block.} producer sends each chunk to the root of its tree, which then propagates it further. To tolerate failures, Turbine applies Forward Error Correction (FEC) with a 32:32 rate, encoding each block into 32 data and 32 recovery chunks, doubling the transmitted data.

\begin{figure*}
\centering
    \begin{subfigure}[c]{0.32\textwidth}
        \includegraphics[width=\textwidth]{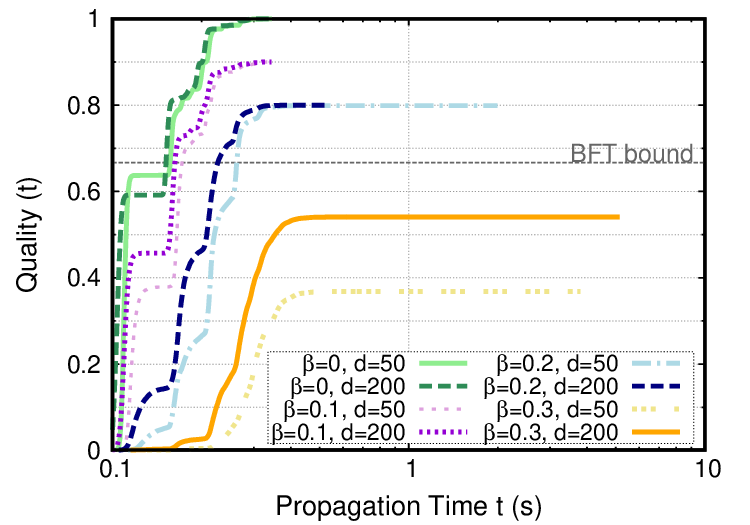}
        \caption{Impact on \namerobust ($\networksize=\num{1000}$)}
        \label{fig:solana-bound-robust}
    \end{subfigure}
    \begin{subfigure}[c]{0.32\textwidth}
        \includegraphics[width=\textwidth]{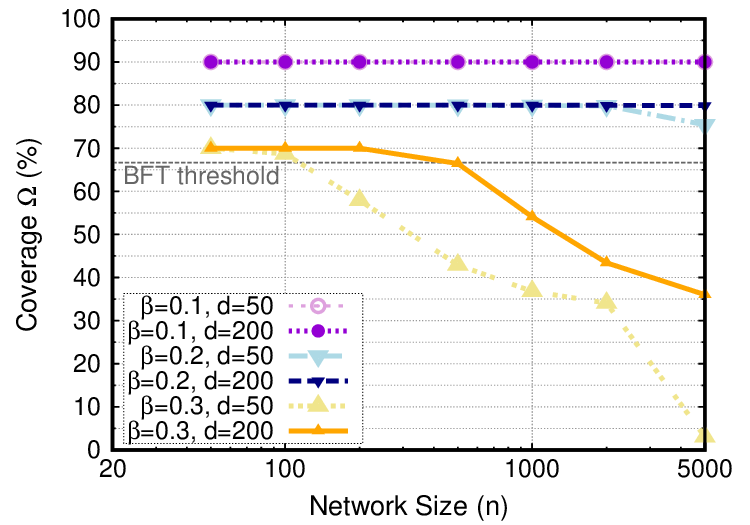}
        \caption{Impact of $\bound$ on coverage}
        \label{fig:solana-bound-coverage}
    \end{subfigure}
    \begin{subfigure}[c]{0.32\textwidth}
        \includegraphics[width=\textwidth]{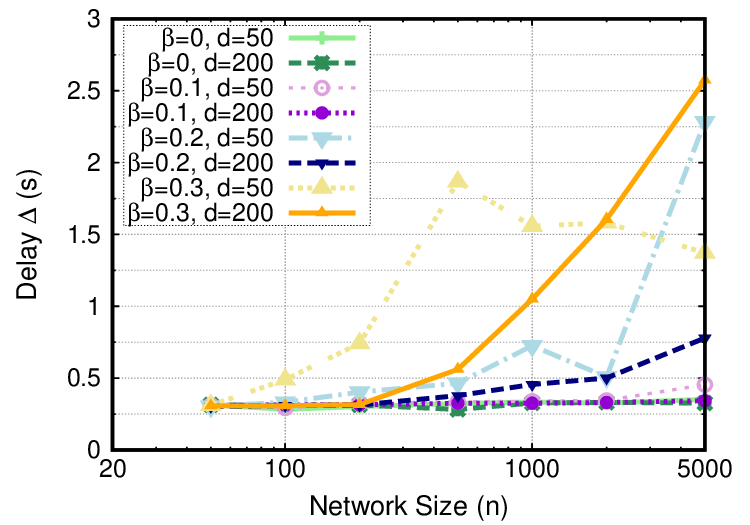}
        \caption{Impact of $\bound$ on delay}
        \label{fig:solana-bound-delay}
    \end{subfigure}
    \caption{Solana (dynamic trees): Impact of $\bound$ on the robustness.}
    \label{fig:solana}
\end{figure*}
Chunk sizes are capped at \num{1232} bytes, derived from the IPv6 MTU and Solana’s QUIC-based transport. We adopt the same chunk size and adjust the FEC rate to 64:64 for our 80 KB messages. Under this setting, a node successfully receives a message once it obtains any 64 chunks, allowing up to $50\%$ chunk loss.

Given that the Solana mainnet has roughly \num{1000} nodes and a fanout of about 200, we simulate both $\nodedegree=200$ and a smaller fanout of $\nodedegree=50$ to capture more conservative and bandwidth-efficient configurations (Figure~\ref{fig:solana}). For $\nodedegree=50$, the dissemination tree has height $2$ when $100 \le \networksize \le 2000$ and increases to $3$ at $\networksize=5000$. For $\nodedegree=200$, the tree height remains $2$ for all $\networksize \ge 500$. 

In our evaluation, we deliberately assume uniform stake distribution across nodes. This setting fits our study: with equal stake, no node is more attractive, whereas any skew concentrates influence and makes it easier for the adversary to corrupt nodes (corrupting high-stake nodes drives coverage to zero).
It is also a fairer basis for comparison as our main aim is to compare the robustness of gossip layers, not consensus mechanisms, and the pairing is sometimes incidental: Bitcoin could in principle run GossipSub, and Solana an ER overlay. Nevertheless, we also validate our findings under Solana's real-world stake distribution, as reported by~\cite{solana_stake_distribution} (see Figure~\ref{fig:solana_stake_ablation} in Appendix~\ref{sec:er_add_plots}). In this setting, coverage is measured as the fraction of total stake held by nodes that received the data. Random corruption (orange in Figure~\ref{fig:solana_stake_ablation}) yields a trend similar to the equal-stake setting (green), whereas \emph{stake-weighted corruption} (in blue in Figure~\ref{fig:solana_stake_ablation}), which prioritizes high-stake nodes, prevents dissemination entirely.

This confirms that our findings are not an artifact of the equal-stake assumption. Rather, equal stake represents a favorable setting for dissemination, since it removes the adversary's ability to exploit stake concentration when selecting nodes to corrupt.

Figure~\ref{fig:solana-bound-coverage} shows that under a weak adversary ($\bound \le 0.2$), coverage remains healthy until the network size reaches $\networksize=\num{5000}$ nodes for $\nodedegree=50$, potentially due to the increase in tree height. In contrast, for $\bound = 0.3$, coverage drops sharply as soon as $\networksize \geq 50$ and approaches $\approx 0\%$ ($\nodedegree=50$), indicating that more than half of the chunks are lost, and $\approx 36\%$ ($\nodedegree=200$) for $\networksize \geq \num{5000}$ (i.e., tree heights $\geq 3$). 

As shown in Figure~\ref{fig:solana-bound-delay}, dynamic trees with chunk-based push achieve consistently low latency when coverage is high. For $\bound=0.1$, dissemination delay stays around $0.3$~s across all network sizes, matching the no-adversary baseline. With $\bound=0.2$ and $80\%$ coverage, dissemination in a \num{5000}-node network takes only $0.75,\mathrm{s}$ for $\nodedegree=200$, compared to at least $2$~s in other configurations. Once coverage degrades under stronger adversaries, latency becomes highly variable.
Figure~\ref{fig:solana-bound-robust} shows a stair-shaped \namerobust curve for $\networksize = \num{1000}$, where the tree height is $2$ for both fanouts. Flat regions correspond to periods in which chunks are propagating, but nodes have not yet received enough chunks to reconstruct the message, while sharp increases occur when many nodes simultaneously cross the reconstruction threshold.

\subheading{Impact of Churn.}
Figure~\ref{fig:solana-churn-coverage} shows that dynamic trees with chunk-based propagation are highly sensitive to churn when $\bound \ge 0.2$. With $30\%$ churn nodes, coverage already falls below the BFT threshold \bftrelbound. Since offline nodes behave, in terms of coverage, equivalently to corrupted nodes, this result is consistent with the static-network findings in Figure~\ref{fig:solana-bound-coverage}, indicating that Solana becomes insecure once the corruption bound exceeds $0.2$.

\subheading{Impact of Network Size.}
With a fixed degree, coverage is affected by network size as the tree grows taller. Although the protocol mitigates hierarchical effects---where nodes closer to the root are more influential---by randomizing the propagation tree for each chunk, Figure~\ref{fig:solana-bound-coverage} still shows a strong correlation between coverage and tree height.
Specifically, for $\nodedegree=50$, the height first increases at $\networksize=100$, leading to insufficient coverage once $\networksize\geq 200$ given $\bound=0.3$, and increases again at $\networksize=\num{5000}$, causing another coverage drop for $\bound=0.2$. Similarly, for $\nodedegree=200$, coverage decreases at $\networksize=500$ when the tree height rises from 1 to 2 (the tree height in Turbine is determined as $h=\lceil log_{\nodedegree}(\networksize) \rceil$).  

\section{Interpretation of Findings}  \label{sec:findings}

In the following, we interpret our findings through the lens of our research questions.

\subsection{RQ1: When are Gossip Protocols Insecure?}

Across all evaluated settings, several consistent trends emerge in coverage and delay under adversarial conditions.

For ER graphs, coverage is largely independent of network size (Figure~\ref{fig:er-strategy-coverage} and Figure~\ref{fig:er-bound-coverage} in Appendix~\ref{sec:er_add_plots}), whereas delay grows with it (Figure~\ref{fig:er-strategy-delay}). Increasing the node degree notably improves robustness for large adversarial bounds $\bound$, but has little impact for small ones. For networks of up to \num{5000} nodes, this gain saturates at approximately $\nodedegree \geq 15$, well above $\ln(\num{5000}) \approx 8.5$. Beyond this point, higher degrees yield no meaningful improvement in coverage and only add bandwidth overhead (Figure~\ref{fig:er-degree-comp}). For production deployments, the saturation point thus marks a practical upper bound on the connectivity worth investing in. Churn reduces coverage additionally under otherwise identical network and adversarial parameters.

Structured grid topologies, in contrast, keep coverage above the bounds in all considered configurations owing to their high connectivity. This makes them best suited to production deployments in which node bandwidth is not the limiting constraint.

We note that GossipSub exhibits insufficient coverage at low transaction rates, even without adversaries. Its peer-scoring mechanism implicitly assumes more than one transaction per second and predictable link latencies: honest nodes are expected to relay a certain number of messages within a short delivery window or incur penalties. Deploying GossipSub in blockchains with lower transaction rates therefore requires carefully tuning parameters such as the score refresh interval, the penalty weight, and the delivery window (see Table~\ref{tb:gossipsub-params}), so as to remain fair to honest nodes while still detecting misbehavior. Moreover, GossipSub is vulnerable to selective dropping: an adversary that forwards some messages but drops others can push coverage below the \bftboundtext, while its score remains largely unaffected. Overall, GossipSub's scoring function depends heavily on network assumptions, which makes it inflexible and leaves the network vulnerable to adversaries with faster connections. Interestingly, its delays are largely insensitive to both network size and $\bound$.

For dynamic trees with double recovery chunks, coverage is ideal ($1 - \bound$) for $\bound = 0.1$. For $\bound = 0.33$, however, it degrades sharply, from $67\%$ at $\networksize = 200$ to at most $15\%$ at $\networksize = \num{5000}$. Chunk-based dissemination with FEC $32{:}32$ over dynamic trees thus cannot guarantee sufficient coverage for any network size or fanout satisfying $\networksize > \nodedegree$ when $\bound = 0.33$. Consequently, if one third of the approximately \num{1000} nodes in Solana's mainnet crash, only about $30\%$ of nodes can reconstruct messages via Turbine, which is insufficient to form a quorum in Solana's consensus protocol.

Finally, churn substantially weakens the robustness of all four gossip protocols, often violating the coverage required for consensus.

\subsection{RQ2: How Do Gossip Protocols Compare?}

The best coverage is achieved by ER graphs with sufficiently large node degree ($\nodedegree \ge 15$), by grid topologies across most settings, by GossipSub with a targeted node degree of $\nodedegree = 8$ when message rate $\arrivalrate$ is high, and in Solana for fully connected networks. 

Overall, Solana's dynamic trees are the least reliable protocol: increasing the network size or decreasing the node degree markedly reduces coverage. GossipSub is highly sensitive to the transaction rate, but at sufficiently high $\arrivalrate$, its coverage matches that of grids and random graphs with only about half the node degree, since its peer scoring preferentially disconnects underperforming peers. Under churn, grids consistently outperform ER graphs, with about $1\%$ higher coverage for $\bound = 0.1$ and churn rate $c = 0.05$, and up to $4.1\%$ higher coverage for $\bound = 0.3$ and $c = 0.5$.

Without an adaptive adversary, ER graphs ($\nodedegree = 15$), GossipSub ($\nodedegree = 8$), dynamic trees ($\nodedegree = 200$), and grids achieve comparable coverage in a \num{1000}-node network. Dynamic trees are the most bandwidth-efficient, as erasure coding keeps duplication low, and transfer about $142\,\mathrm{MB}$ in total (Table~\ref{tab:bandwidth}). They are followed by GossipSub ($337\,\mathrm{MB}$) and ER graphs ($685\,\mathrm{MB}$). Grids are the least efficient, requiring about $2.97\,\mathrm{GB}$ to disseminate a single $80\,\mathrm{KB}$ block. At Polkadot's rate of one block every $6$ seconds, this amounts to $41.75\,\mathrm{GB}$ per node per day.

At $\bound = 0.33$, with corrupted nodes not relaying messages, total bandwidth drops to about $226\,\mathrm{MB}$ for GossipSub, $515\,\mathrm{MB}$ for ER graphs, and $2.3\,\mathrm{GB}$ for grids, i.e., $67\%$, $75\%$, and $77\%$ of the benign volume, respectively. We exclude dynamic trees here, since they no longer provide sufficient guarantees at this bound. The higher relative volume of ER graphs and grids reflects data lost along paths through corrupted nodes, whereas GossipSub actively excludes poorly performing peers from dissemination paths. This efficiency, however, comes at the cost of sensitivity to parameter tuning: if the expected message rate or network latency is misestimated, or if an adversary selectively drops messages, GossipSub's coverage can degrade severely.

\begin{table}
    \scalebox{0.95}{
    \begin{tabular}{|c|c|c|c|c|} 
        \hline
        $\bound$ & \makecell{Random Graphs\\($\nodedegree=15$)}& \makecell{Grid\\(Polkadot)}& \makecell{GossipSub\\($\nodedegree=8$)}& \makecell{Dynamic Trees\\(Solana $\nodedegree=200$)}\\
        \hline
        \hline
        0& 685& 2969& 337&142\\\hline
        0.1& 645& 2798& 304&120\\\hline
        0.2& 595& 2599& 269&100\\\hline
        0.33& 515& 2299& 225&--\\\hline
        0.5& 387& 1835& 167& -- \\\hline    
    \end{tabular}
    }
    \caption{Transmitted data (MB) to gossip an 80~KB message in a \num{1000}-node network (results for dynamic trees with $\bound \ge 0.33$ are omitted due to insufficient coverage).}
    \label{tab:bandwidth}
\end{table}

Taken together, these findings indicate that none of the modern gossip protocols we study achieves a markedly better tradeoff between robustness, performance, and bandwidth than a simple Erd\H{o}s--R\'enyi (ER) random graph, a structure whose properties are well understood. The added protocol complexity (e.g., of Dynamic Trees and GossipSub) does not necessarily translate into added security. Random graphs instantiated with a sufficient degree ($d \geq 15$) thus emerge as a practical choice for production deployments, offering a strong balance between robustness and performance while remaining simple to analyze and implement. 
\section{Related Work}

Almost all consensus protocols, including traditional BFT protocols and Proof-of-Work or Proof-of-Stake protocols, are designed under the assumption that message delivery is eventually guaranteed among all honest available nodes. Our work shows that existing gossip protocols~\cite{karp2000randomized,gossipsub,polkadot-grid,solana-turbine} fail to provide that guarantee in practice, and we introduce adversarial strategies that exploit this observation and are not captured by existing models.

\subheading{Efficient Gossip Protocols for Blockchains.}
A substantial body of work improves the efficiency of gossip-based block dissemination by optimizing bandwidth, latency, and routing~\cite{compactblock, erlay, perigee, mercury, gossipsub}. Bitcoin’s compact blocks~\cite{compactblock} reduce bandwidth by sending short transaction identifiers instead of full blocks, at the cost of an extra round-trip for reconstruction. Erlay~\cite{erlay} further reduces overhead via set reconciliation, trading bandwidth efficiency for higher latency.

Other systems optimize routing rather than message format. BlockP2P~\cite{blockp2p} exploits geographic locality to accelerate intra-cluster propagation, but increases vulnerability to partitioning if adversaries target inter-cluster links. Mercury~\cite{mercury} uses virtual coordinate systems to estimate efficient paths and applies a high initial fanout to speed up early dissemination, though its security degrades if coordinates are spoofed. Perigee~\cite{perigee} constructs an optimized routing graph based on peer scoring and observed block arrival times; however, such adaptive routing is susceptible to eclipse attacks by well-connected adversaries.
Kumar \etal~\cite{gossipsub-analysis} formally analyze GossipSub's \emph{scoring function} and show that its correctness hinges on the choice of weights, which corroborates our findings. We significantly extend their analysis by measuring GossipSub's effective coverage alongside three other gossip protocols, enabling a direct comparison of their security.

\subheading{Robust Gossip Protocols.}
Despite their importance, the robustness of gossip protocols under adversarial conditions has received limited attention. Early work by Karp \etal~\cite{karp2000randomized} and Kermarrec \etal~\cite{kermarrec2003probabilistic} analyzes rumor spreading on dynamic or random graphs under benign failures, characterizing the trade-off between fanout and reliability while largely abstracting away Byzantine behavior and adversarial membership manipulation.
To address Byzantine faults, Corretti \etal~\cite{scuttlebutt} propose a gossip layer for proof-of-stake blockchains that samples neighbors using VRFs and stake-weighted connections, providing Sybil resistance but remaining vulnerable to eclipse attacks under sufficient adversarial stake. More recently, Albrecht \etal~\cite{albrecht2024larger} studied the impact of gossip behavior on NS-blockchains, but only consider a static adversary that randomly corrupts edges, and do not consider adaptive or targeted network attacks. 

\subheading{Network Attacks on Blockchains.}
A complementary line of work examines attacks on blockchain network layers. Eclipse attacks~\cite{eclipse-bitcoin, eclipse-ethereum} exploit weaknesses in peer discovery to isolate victims with adversarial peers, enabling message filtering or delay; Gervais \etal~\cite{tempering-delivery} show that this can substantially slow Bitcoin’s propagation without full partitioning.
Routing-level attacks further exacerbate these threats. 
Apostolaki \etal~\cite{routing-attack} demonstrate that BGP hijacks can partition Bitcoin or severely delay propagation, while stealthier variants~\cite{erebus} poison routing paths via spoofed IPs. To counter routing-level attacks, Apostolaki \etal~\cite{sabre} proposed SABRE, which mitigates such attacks by leveraging a coexisting relay network with secure channels. However, this approach requires trust in the relay nodes.
In this paper, we show that an adversary can simply exploit the \emph{topology} adaptively to achieve partition without exploiting vulnerabilities in node discovery protocols, and without requiring additional resources such as compromised ASs or IPs.
\section{Conclusion}

In this paper, we analyzed the robustness of the gossip protocols used in popular decentralized platforms. Our analysis and extensive measurements reveal a notable gap between the guarantees classical BFT protocols require and those these gossip protocols provide against realistic network adversaries.

Specifically, we evaluated random graphs, Polkadot's grid topology, Ethereum's GossipSub, and Solana's chunk- and tree-based protocols. Polkadot is the most robust under adaptive attacks, combining consistently high coverage with stable delays, albeit at a very high bandwidth cost. Random graphs achieve similar robustness with a sufficiently high degree ($\nodedegree \geq 15$). In contrast, Bitcoin's degree of 8, which coincides with the recommended connectivity for random graphs~\cite{Erdos_Renyi}, and Solana's dynamic trees prove unreliable under attack, particularly at scale. GossipSub yields more variable performance and is highly sensitive to parameter tuning. Overall, despite their simplicity, random graphs with a sufficient degree emerge as a practical choice for decentralized deployments, offering a strong tradeoff between robustness and performance.

To ensure reproducibility, the simulator code and the data for all experiments are available at~\cite{repo-simulator}.

\ifCLASSOPTIONcompsoc
  \section*{Acknowledgments}
\else
  \section*{Acknowledgment}
\fi

\noindent The authors thank the reviewers for their helpful feedback.
This work has been partly funded by the Deutsche Forschungsgemeinschaft (DFG, German Research Foundation) under Germany's Excellence Strategy -- EXC 2092 CASA -- 390781972.

\bibliographystyle{ieeetr}
\bibliography{refs}

\appendices
\section{Centrality Metrics} \label{sec:centrality_metrics}

Our analysis (cf.~Sections~\ref{sec:Breaking-BFT-Bound} and~\ref{sec:evaluation}) uses the four most common centrality metrics:

\begin{description}[leftmargin=0.2cm]
    \item[Degree centrality] captures a node's one-hop connectivity as its degree normalized by the maximum possible degree, $\frac{\deg(\node)}{\networksize-1}$.
    \item[Eigenvector centrality] accounts not only for a node's links but also for the centrality of its neighbors. It is given by the corresponding entry of the principal eigenvector $x$ of the adjacency matrix $A$, i.e., $Ax = \lambda x$.
    \item[Betweenness centrality] measures how often a node acts as a bridge, which is especially relevant in clustered networks. It is defined as $\sum_{\node_1,\node_2 \in \nodeset\setminus\{\node\}} \frac{\sigma_{\node_1\node_2}(\node)}{\sigma_{\node_1\node_2}}$, where $\sigma_{\node_1\node_2}$ is the number of shortest paths from $\node_1$ to $\node_2$ and $\sigma_{\node_1\node_2}(\node)$ the number of those passing through $\node$.
    \item[Closeness centrality] reflects a node's proximity to the rest of the network as $\frac{\networksize-1}{\sum_{\node' \neq \node} \mathrm{dist}(\node,\node')}$, where $\mathrm{dist}(\node,\node')$ is the shortest-path length between $\node$ and $\node'$.
\end{description}

\section{Simulator Validation} \label{sec:validation}

As mentioned earlier, we cannot conduct our evaluation on live networks directly. Doing so would disrupt real users and put funds at risk, and it would also be infeasible, since we cannot control parameters such as the size of an existing network. One of the few workable means to validate our results is therefore to show that our simulator faithfully reproduces the behavior of the real networks in the benign setting. To this end, we compare the message-arrival-time distributions observed by our simulated nodes with those observed by unmodified validator nodes in each of the blockchain networks we consider (as the availability of such measurements varies widely across blockchains, we obtain them through network-specific means). For each network, we compare both the CDF and the PDF of message latencies and check whether the simulated distributions match those observed in practice.

\begin{figure}[tbp]
    \centering
    \begin{subfigure}[c]{0.49\linewidth}
        \includegraphics[width=\textwidth]{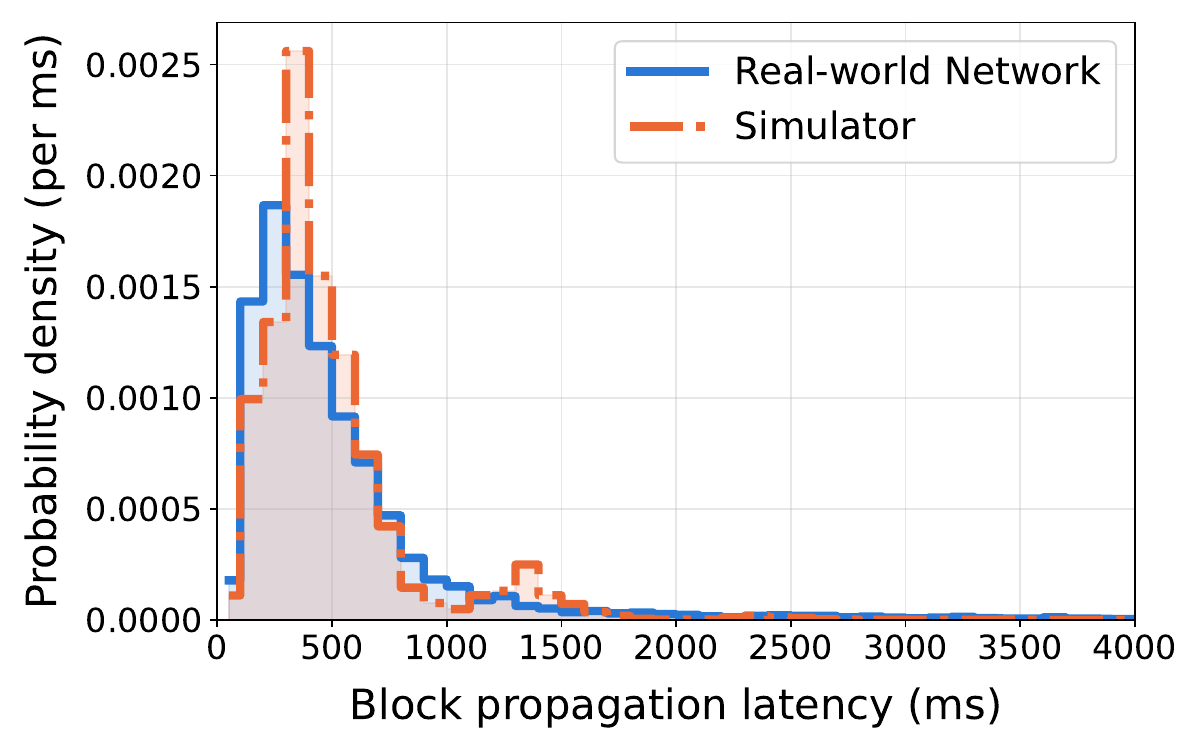}
        \caption{Bitcoin PDF}
        \label{fig:Bitcoin_validation_PDF}
    \end{subfigure}
    \begin{subfigure}[c]{0.49\linewidth}
        \includegraphics[width=\textwidth]{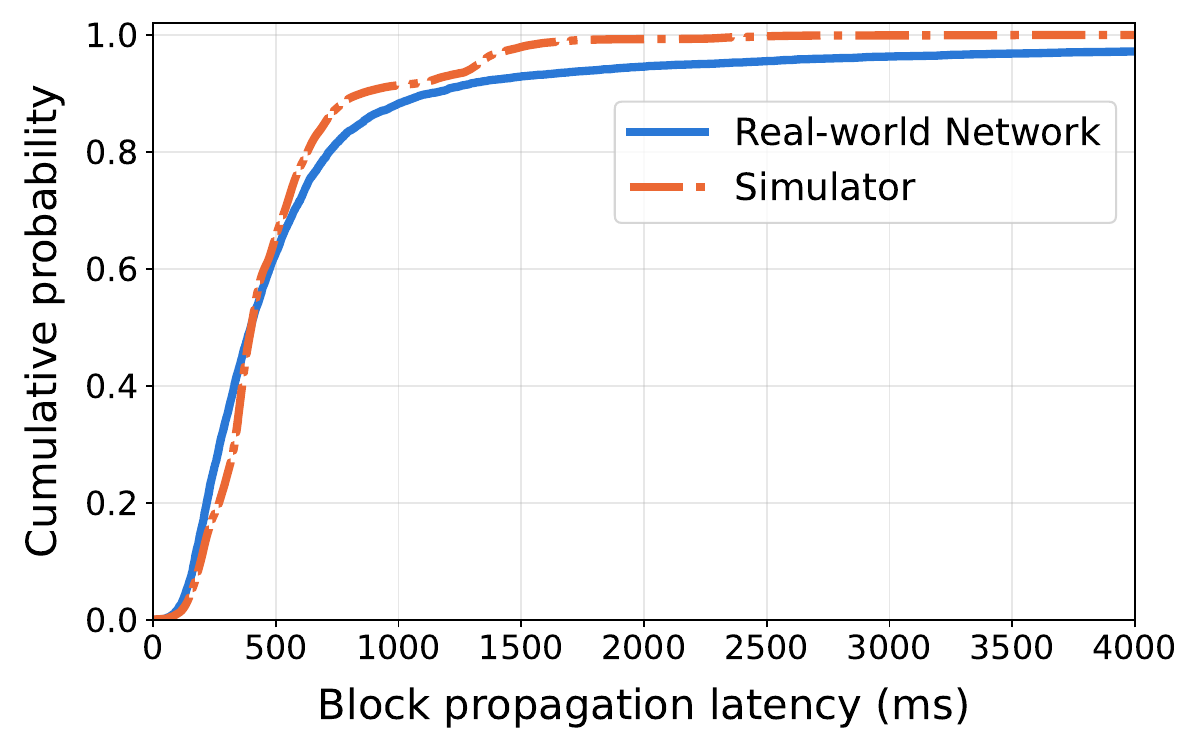}
        \caption{Bitcoin CDF}
        \label{fig:Bitcoin_validation_CDF}
    \end{subfigure}
    \begin{subfigure}[c]{0.49\linewidth}
        \includegraphics[width=\textwidth]{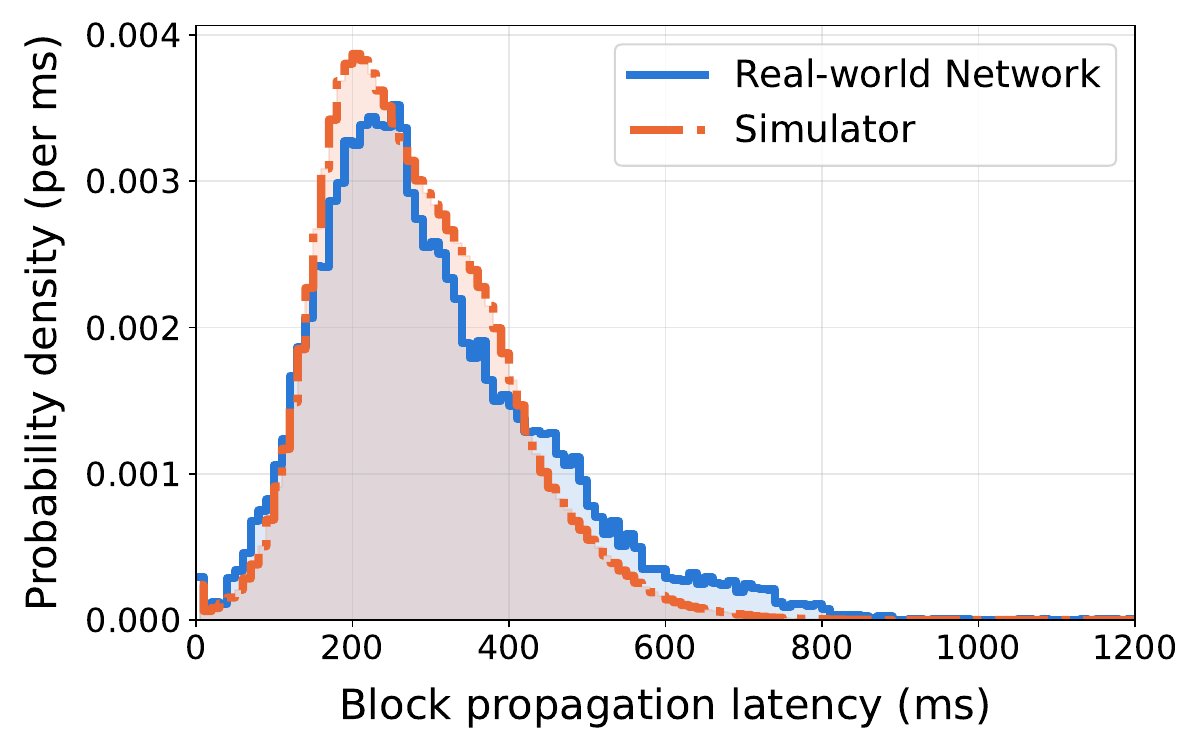}
        \caption{Polkadot PDF}
        \label{fig:Polkadot_validation_PDF}
    \end{subfigure}
    \begin{subfigure}[c]{0.49\linewidth}
        \includegraphics[width=\textwidth]{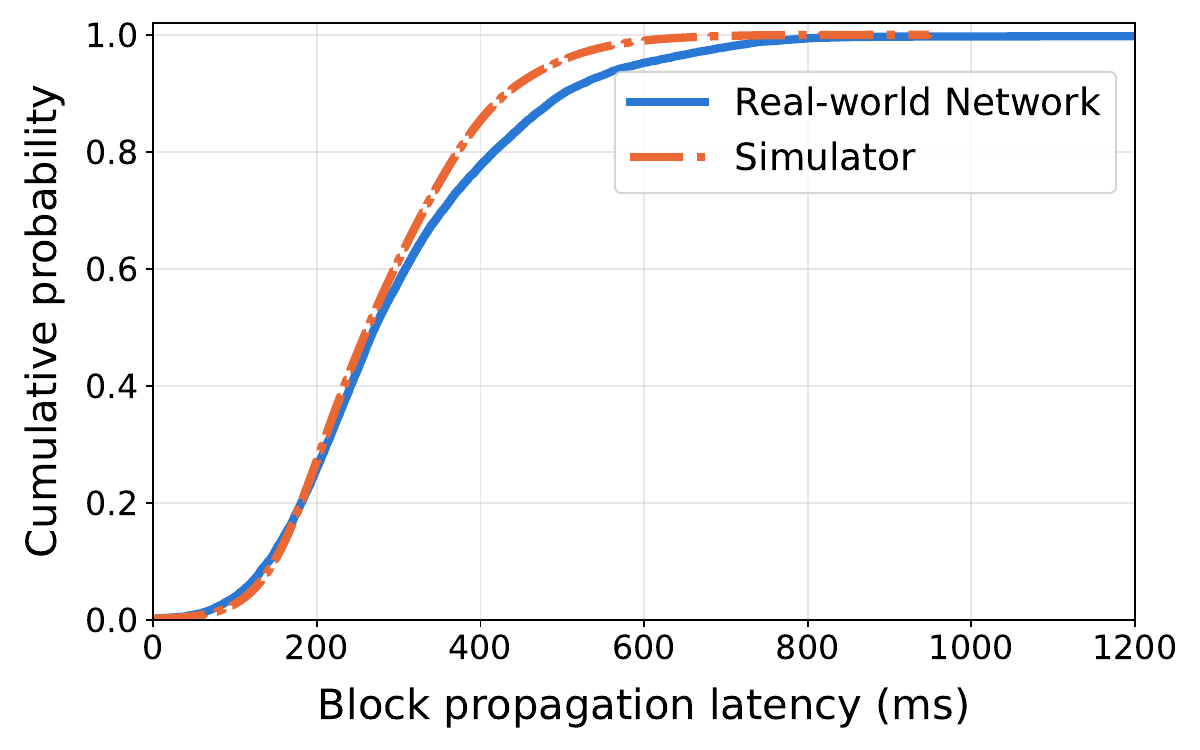}
        \caption{Polkadot CDF}
        \label{fig:Polkadot_validation_CDF}
    \end{subfigure}
    \centering
    \begin{subfigure}[c]{0.49\linewidth}
        \includegraphics[width=\textwidth]{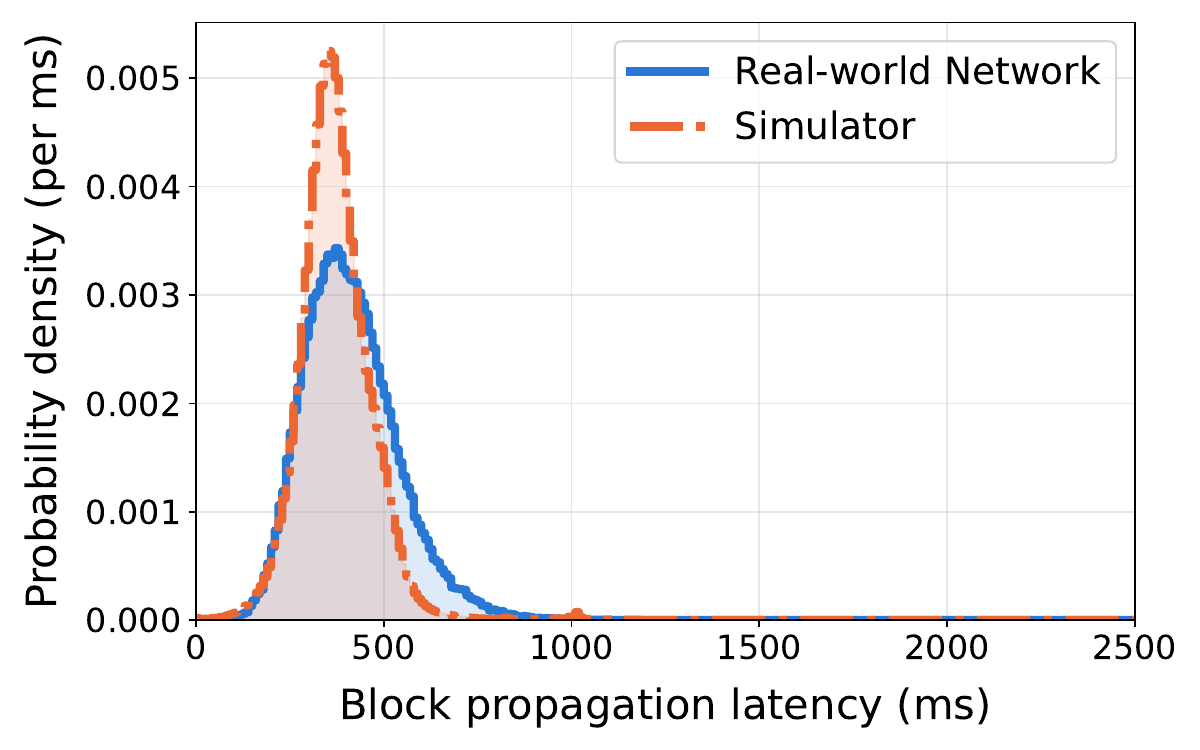}
        \caption{Ethereum PDF}
        \label{fig:Ethereum_validation_PDF}
    \end{subfigure}
    \begin{subfigure}[c]{0.49\linewidth}
        \includegraphics[width=\textwidth]{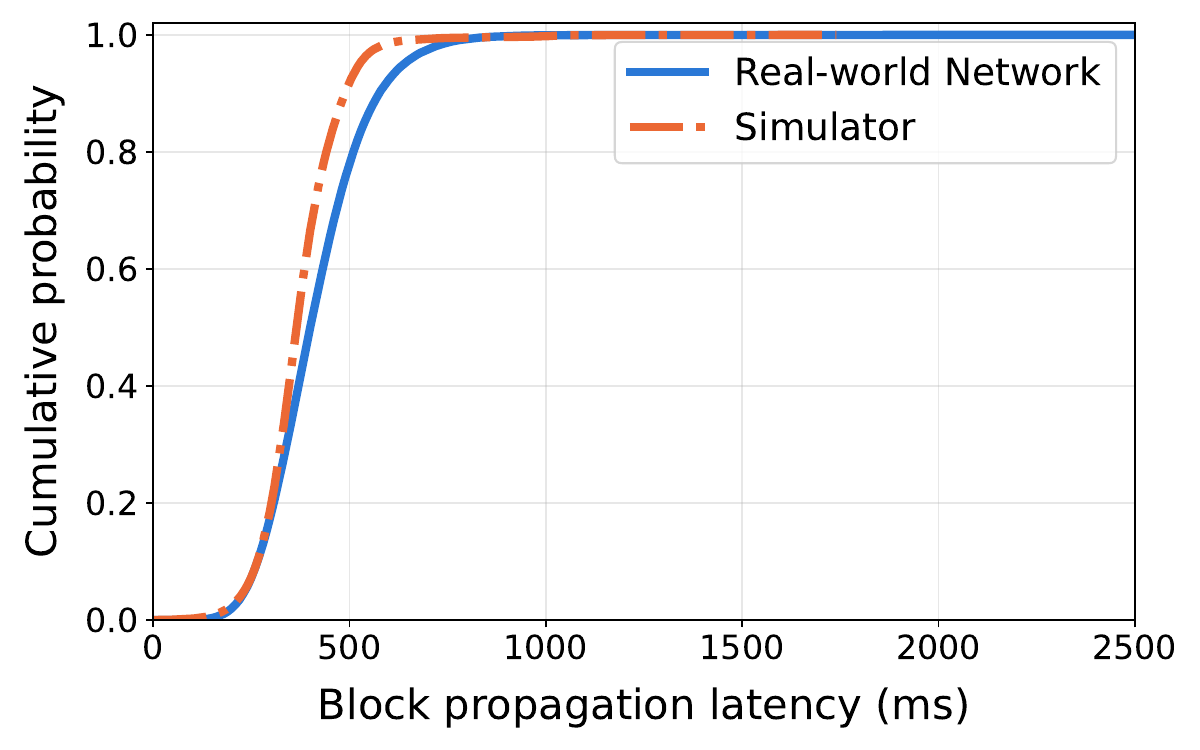}
        \caption{Ethereum CDF}
        \label{fig:Ethereum_validation_CDF}
    \end{subfigure}
    \centering
    \begin{subfigure}[c]{0.49\linewidth}
        \includegraphics[width=\textwidth]{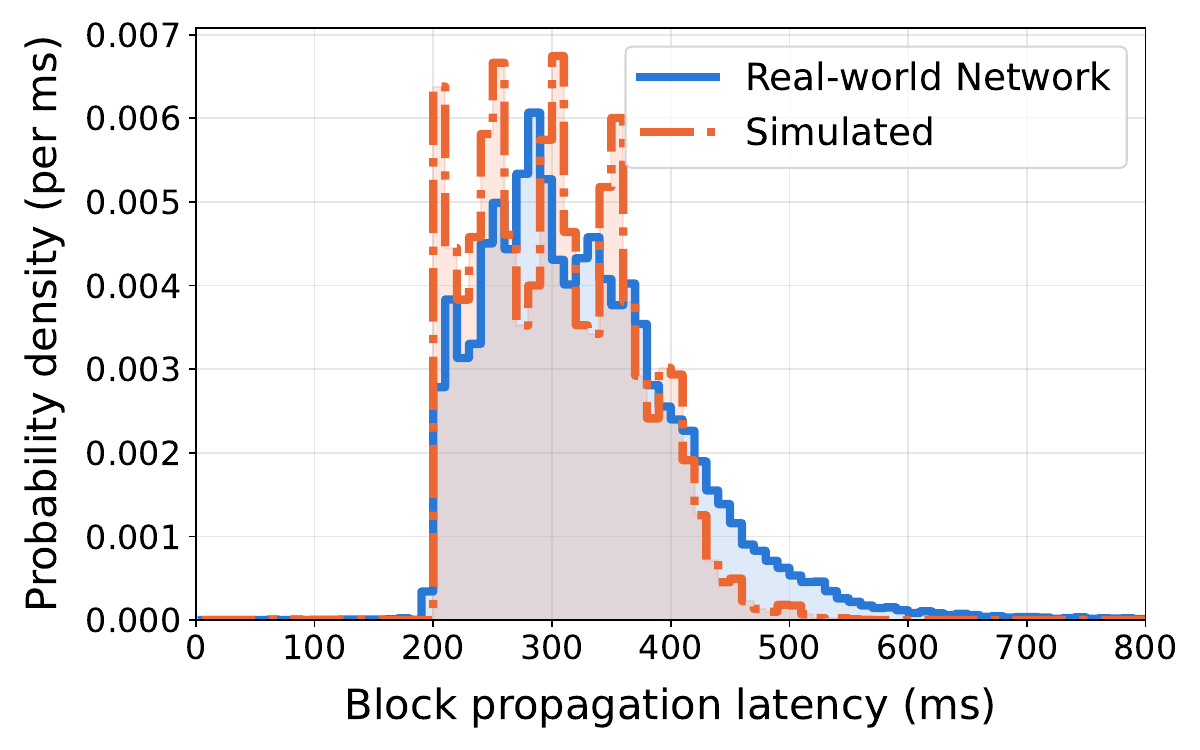}
        \caption{Solana PDF}
        \label{fig:Solana_validation_PDF}
    \end{subfigure}
    \begin{subfigure}[c]{0.49\linewidth}
        \includegraphics[width=\textwidth]{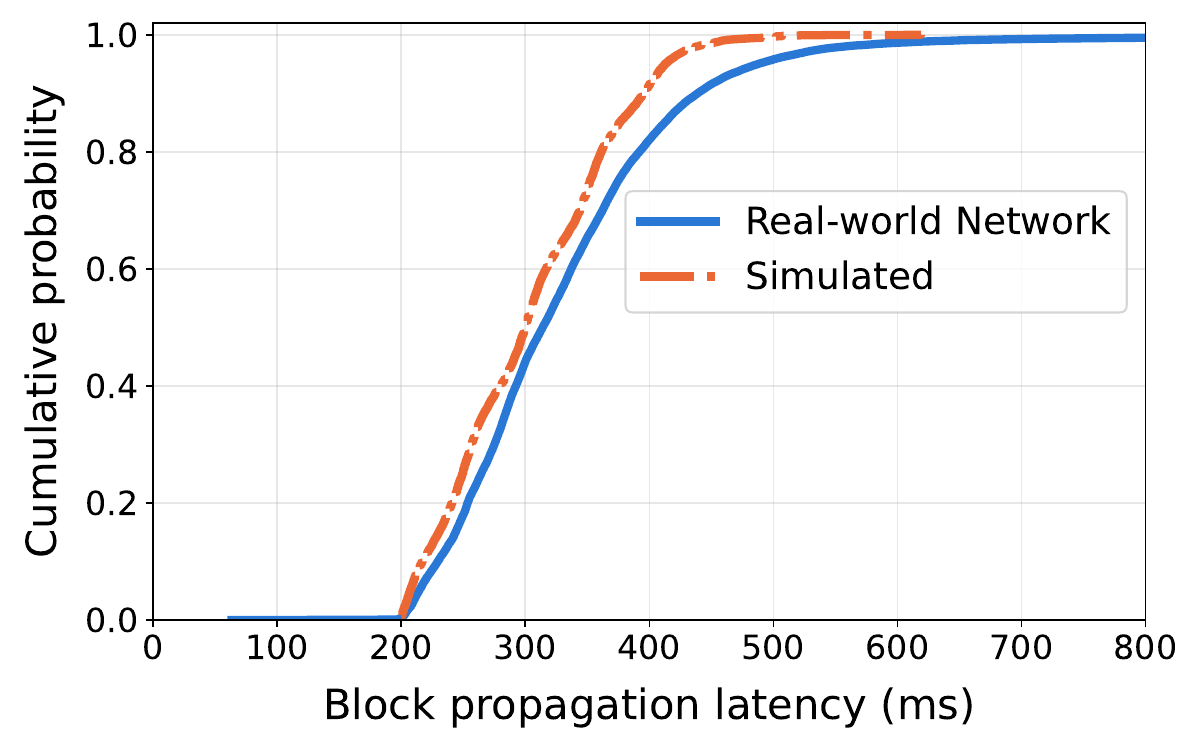}
        \caption{Solana CDF}
        \label{fig:Solana_validation_CDF}
    \end{subfigure}
    \caption{
        CDF and PDF of block propagation latencies observed in Bitcoin, Polkadot, Ethereum, and Solana vs.\ those produced by our simulator. 
    }
    \label{fig:Simulator_validation}
\end{figure}

\subheading{\underline{Bitcoin}.}
For Bitcoin, we use block propagation delays publicly reported in~\cite{neudecker-atc16} and derived from block announcements by unmodified Bitcoin nodes. Among the collected network statistics, we use the time between the first reception of an \texttt{INV} message announcing a new block and its subsequent receptions from other peers, retrieved on 2022-09-11. To faithfully compare with our simulator, we simulate \num{10000} nodes with a message size of \SI{1.6}{\mega\byte}, matching the Bitcoin network configuration.

Figure~\ref{fig:Bitcoin_validation_PDF} and Figure~\ref{fig:Bitcoin_validation_CDF} show the PDF and CDF, respectively, of the latency distribution produced by our simulator and of the block propagation distribution measured from Bitcoin nodes. The two distributions are highly similar, with a Pearson correlation coefficient of $\rho = 0.93$.

\subheading{\underline{Polkadot}.}
We measure block propagation delays using the public Polkadot telemetry service, a collector to which Polkadot nodes stream operational events such as block imports in real time. Since the stock Polkadot client reports to this service by default, telemetry offers a passive view of propagation as experienced by unmodified nodes on the live network, without requiring us to deploy or instrument any nodes ourselves. We subscribed to the telemetry feed (\texttt{wss://feed.telemetry.polkadot.io/feed}) for the Polkadot chain, identified by its genesis hash \texttt{0x91b171bb\ldots}, and recorded every block-import report over a 1-hour window on 2026-09-21. Each report contains a propagation time computed by the collector as the interval between the first report of a block in the network and the reception of that block by each node.

We retain only nodes that publish a validator stash address and discard reports with a propagation time of zero, which correspond to the node that first announced the block, as well as reports exceeding \SI{6}{\second}, which stem from nodes that are resyncing or stalled. Since telemetry participation is self-selected and over-represents nodes hosted in European infrastructure relative to the actual validator set, we compare against a simulator configuration with validators placed predominantly in Europe~\cite{polkadot-region-dist}, with 600 nodes and an 80~KB message size. This way, the simulated population mirrors the measured sample's scale and geographic composition.
 
For Polkadot, the block-arrival-time distribution in our simulator closely matches that measured from unmodified validator nodes, for the PDF (Figure~\ref{fig:Polkadot_validation_PDF}) and the CDF (Figure~\ref{fig:Polkadot_validation_CDF}), with a Pearson correlation coefficient of $\rho = 0.99$.

\subheading{\underline{Ethereum's GossipSub}.}
For Ethereum, we obtain block arrival times from Xatu~\cite{xatu-data}, a public collection of beacon-chain events released by ethPandaOps~\cite{ethpandaops}. Contributors run a Xatu sentry sidecar alongside their unmodified beacon node, which records the node's event stream and publishes it as daily Parquet files. We use the mainnet \verb+libp2p_gossipsub_beacon_block+ table from 2026-09-01 to 2026-09-28, whose field \verb+propagation_slot_start_diff+ gives the time in milliseconds between the start of a block's slot and its observation at the P2P layer. Since this interval also includes block preparation time, which typically exceeds one second on average~\cite{attestation-timing,probelab}, we instead reference each block to its earliest observation across all sentries. We remove repeated events for the same block, e.g., caused by reorgs or client restarts, and discard observations exceeding \SI{12}{\second}. We then compute the time from the block's earliest observation until 95\% of its observing sentries have reported it. As \verb+libp2p_+ events reflect only the network layer, we further retain only blocks that were validated and imported by the consensus layer, as recorded in \verb+beacon_api_eth_v1_events_block+. 
Since the observing sentries are concentrated predominantly in Europe, we adjust the regional distribution of our simulator accordingly and simulate a network of \num{6000} nodes (see consensus-layer validators in~\cite{ethereum-region-dist}). with a message size of \SI{180}{\kilo\byte}. The simulated and observed distributions closely match (Figures~\ref{fig:Ethereum_validation_PDF} and~\ref{fig:Ethereum_validation_CDF}), with a Pearson correlation coefficient of $\rho = 0.95$.

\begin{table}[tb]
    \centering
    \footnotesize
    \setlength{\aboverulesep}{0pt}
    \setlength{\belowrulesep}{0pt}
    \renewcommand{\arraystretch}{1.15} 
    \scalebox{0.95}{
    \begin{tabular}{l|llll}
        \toprule
        $P_1$& $w_1$& 0.0027& $cap_1$& 3600\\ 
        $P_2$& $w_2$& 0.664& $cap_2$& 1500\\ 
        & $decay_2$& 0.9916& & \\ 
        $P_{3a}$& $w_{3a}$& -0.25& $cap_3$& 2 \\ 
        & $threshold_3$& 10& $decay_{3a}$& 0.997\\
        & $\mathsf{ActivationWin_3}$& 1 min & $\mathsf{DeliveryWin_3}$ &5 ms \\ 
        $P_{3b}$& $w_{3b}$& -0.25& $decay_{3b}$& 0.997\\ 
        Degree & $\mathsf{D}$& 8& $\mathsf{D_{lazy}}$& 12\\ 
        & $\mathsf{D_{low}}$& 6& $\mathsf{D_{high}}$& 12\\
        Interval & $\mathsf{HeartbeatInterval}$& 1 sec& $\mathsf{DecayInterval}$&1 sec\\ 
        \bottomrule
    \end{tabular}
    }
    \caption{GossipSub simulation parameters}
    \label{tb:gossipsub-params}
\end{table}

\subheading{\underline{Solana's Dynamic Trees}.}
We obtain observed metrics from Solana's public cluster telemetry~\cite{solana-metrics}, to which unmodified validator nodes can opt in to report timestamped data points for internal events such as block production stages and shred propagation. Since reporting is built into the client and requires no changes to the node software, these metrics reflect propagation as experienced by nodes running on the live network. The reports are collected in a shared InfluxDB instance that backs the public dashboards at \texttt{metrics.solana.com}. From this database, we extract the \verb+fetch_shred_insert_is_full+ metric, which records, for each slot, the time between a node's receipt of the first and the last shred.

Unlike the other systems we consider, Solana does not disseminate a block as a single unit. Instead, the leader streams data continuously throughout its slot. With a slot duration of \SI{250}{\milli\second} and data emitted every \SI{50}{\milli\second}, the first shred a node receives belongs to the batch emitted at offset \SI{0}{\milli\second}, and the last shred to the batch emitted at offset \SI{200}{\milli\second}. To reproduce this, we configure our simulated chunk- and tree-based protocol to emit a 100~KB message every \SI{50}{\milli\second} in a 700-node network. We then derive the slot-level distribution from the per-message latency distribution: we shift the latter by each emission offset from \SI{0}{\milli\second} to \SI{200}{\milli\second}, sum the shifted distributions, and truncate all mass below \SI{200}{\milli\second}. 

Our validation results show that the simulator closely matches the measured real-world data for both the PDF (Figure~\ref{fig:Solana_validation_PDF}) and the CDF (Figure~\ref{fig:Solana_validation_CDF}), with a Pearson correlation coefficient of $\rho = 0.94$.

\section{Peer Scoring in GossipSub}\label{ap:peerscore}
GossipSub's security rests on its peer scoring function. Each node scores each of its mesh peers based on live statistics as $score = \sum_{k=1}^{6} w_k P_k$, where $P_k$ is a scoring indicator and $w_k$ its weight ($P_1$--$P_4$ are topic-specific in the original design; we assume a single topic). Indicators are either capped or decay at every $\mathsf{DecayInterval}$, in which case they are rounded down to zero once below $\mathsf{DecayToZero}$. Since we do not simulate invalid messages, we ignore $P_4$ (invalid messages) and $P_5$--$P_6$ (application-specific scores, e.g., IP co-location penalties) and focus on the following indicators:

\begin{description}[leftmargin=0.2cm]
    \item[$P_1$ (time in mesh)] Time the peer has spent in the mesh, capped at $cap_1$.
    \item[$P_2$ (first message deliveries)] Number of messages the peer delivered first, capped at $cap_2$ and decaying at rate $decay_2$.
    \item[$P_{3a}$ (mesh message delivery rate)] Penalizes slow relaying. Once the peer has been in the mesh for $\mathsf{ActivationWin_3}$, messages (including duplicates) it delivers within $\mathsf{DeliveryWin_3}$ are counted, up to $cap_3$ and decaying at rate $decay_3$. If the count falls below $\mathsf{Threshold_3}$, then $P_{3a} = (\mathsf{Threshold_3} - count)^2$ with weight $w_{3a} < 0$; otherwise, $P_{3a} = 0$. The short delivery window prevents replay-based attacks.
    \item[$P_{3b}$ (mesh message delivery failures)] Number of failed deliveries in lazy-pull mode, with weight $w_{3b} < 0$, decaying over time.
\end{description}

Scores are updated at every $\mathsf{DecayInterval}$, and peers with negative scores are pruned. Upon pruning, $P_1$ and $P_2$ are reset, whereas $P_{3b}$ is retained and incremented by any active $P_{3a}$ penalty, discouraging the re-grafting of poorly performing peers. Consequently, only two behaviors are rewarded: remaining in the mesh for a long time and delivering messages promptly.

\section{Additional Plots} \label{sec:er_add_plots}

\begin{figure}[htp]
    \centering
    \begin{subfigure}[c]{0.49\linewidth}
        \includegraphics[width=\textwidth]{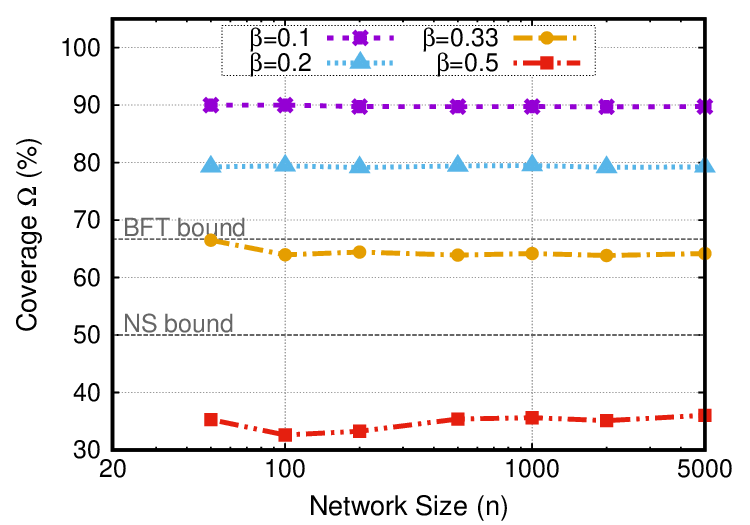}
        \caption{Impact of $\bound$ on coverage}
        \label{fig:er-bound-coverage}
    \end{subfigure}
    \begin{subfigure}[c]{0.49\linewidth}
        \includegraphics[width=\textwidth]{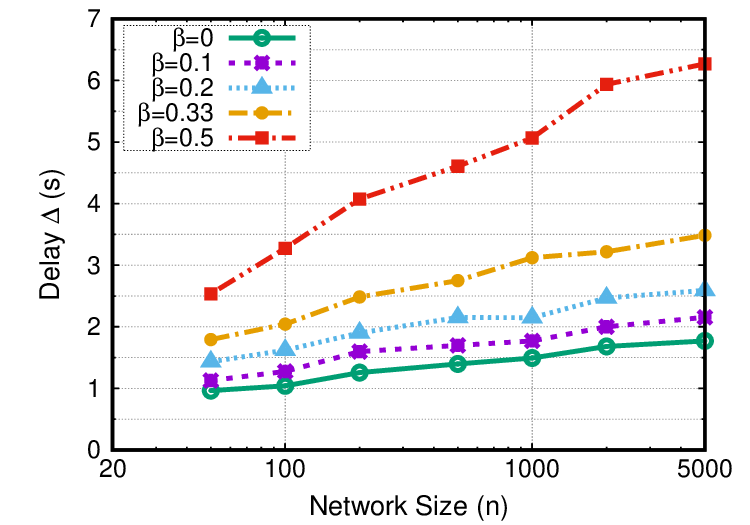}
        \caption{Impact of $\bound$ on delay }
        \label{fig:er-bound-delay}
    \end{subfigure}
    \caption{ER graphs: Impact of $\bound$ (degree-based greedy adversary) on coverage and delay.}
    \label{fig:er-bound-comp}
\end{figure}

\begin{figure}[htp]
    \centering
    \begin{subfigure}[c]{0.49\linewidth}
        \includegraphics[width=\textwidth]{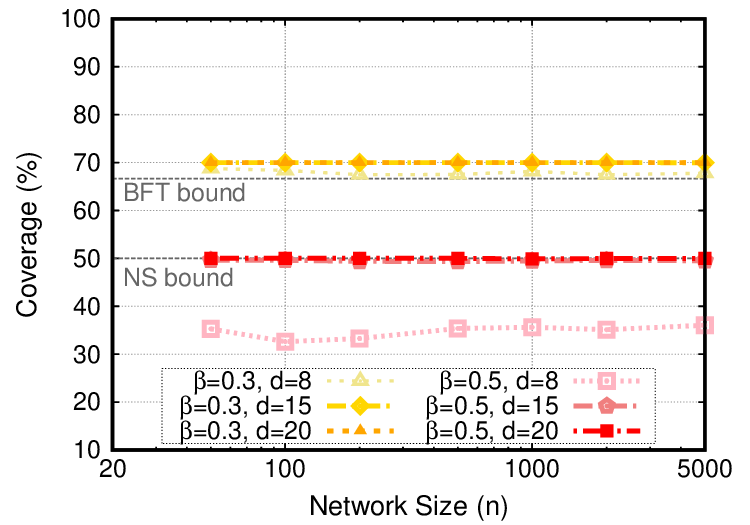}
        \caption{Impact on coverage}
        \label{fig:er-degree-coverage}
    \end{subfigure}
    \begin{subfigure}[c]{0.49\linewidth}
        \includegraphics[width=\textwidth]{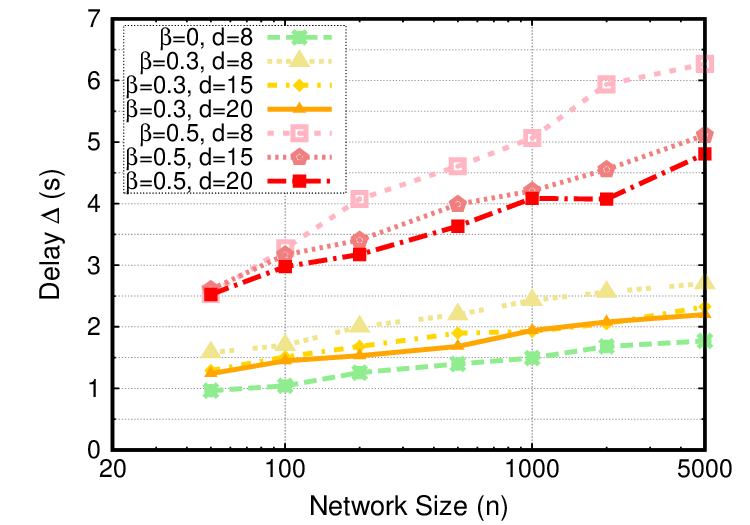}
        \caption{Impact on delay }
        \label{fig:er-degree-delay}
    \end{subfigure}
    \caption{ER graphs: Impact of $\nodedegree$ and $\bound$ (degree-based greedy attacker) on the coverage and delay.} 
    \label{fig:er-degree-comp}
\end{figure}

\begin{figure}[h!]
    \centering
    \begin{subfigure}[c]{0.49\linewidth}
        \includegraphics[width=\textwidth]{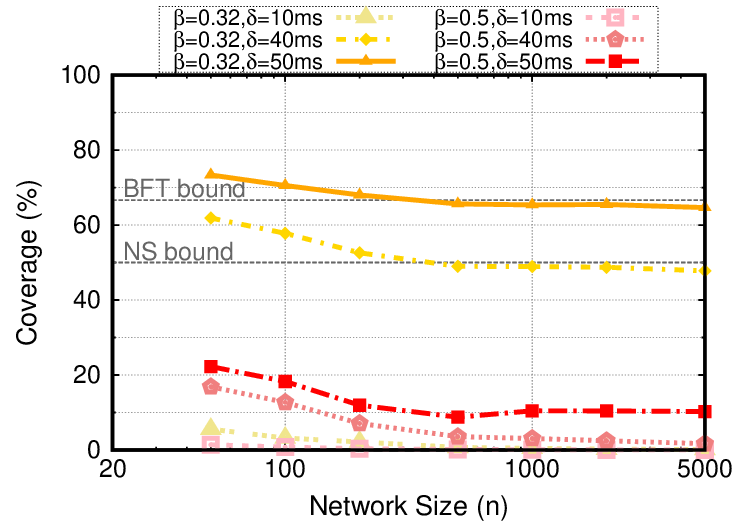}
        \caption{Impact of $\corruptdelay$ on coverage}
        \label{fig:er-adapt-coverage}
    \end{subfigure}
    \begin{subfigure}[c]{0.49\linewidth}
        \includegraphics[width=\textwidth]{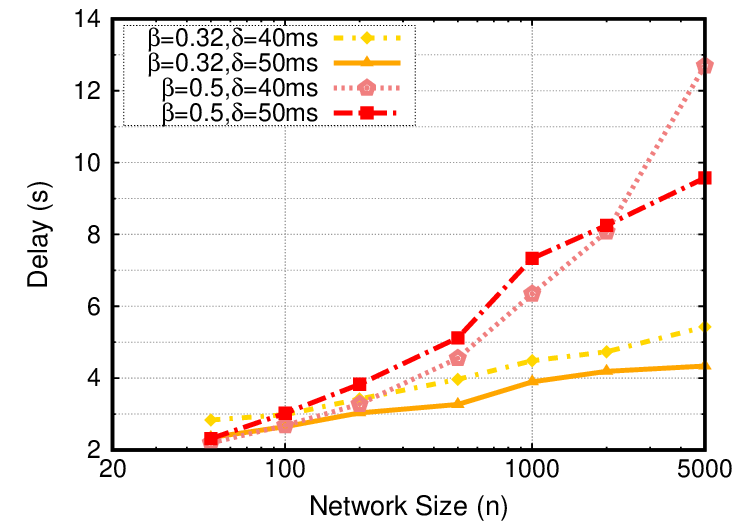}
        \caption{Impact of $\corruptdelay$ on delay}
        \label{fig:er-adapt-delay}
    \end{subfigure}
    \caption{ER graphs: Impact of corruption delays of adaptive adversaries ($\nodedegree=8$).}
    \label{fig:er-adapt-cmp}
\end{figure}

\begin{figure}[h!]
\centering
\renewcommand{\arraystretch}{1.3} 
    \begin{minipage}[t]{0.45\linewidth}
        \vspace{0pt} 
        \centering
        \includegraphics[width=\linewidth]{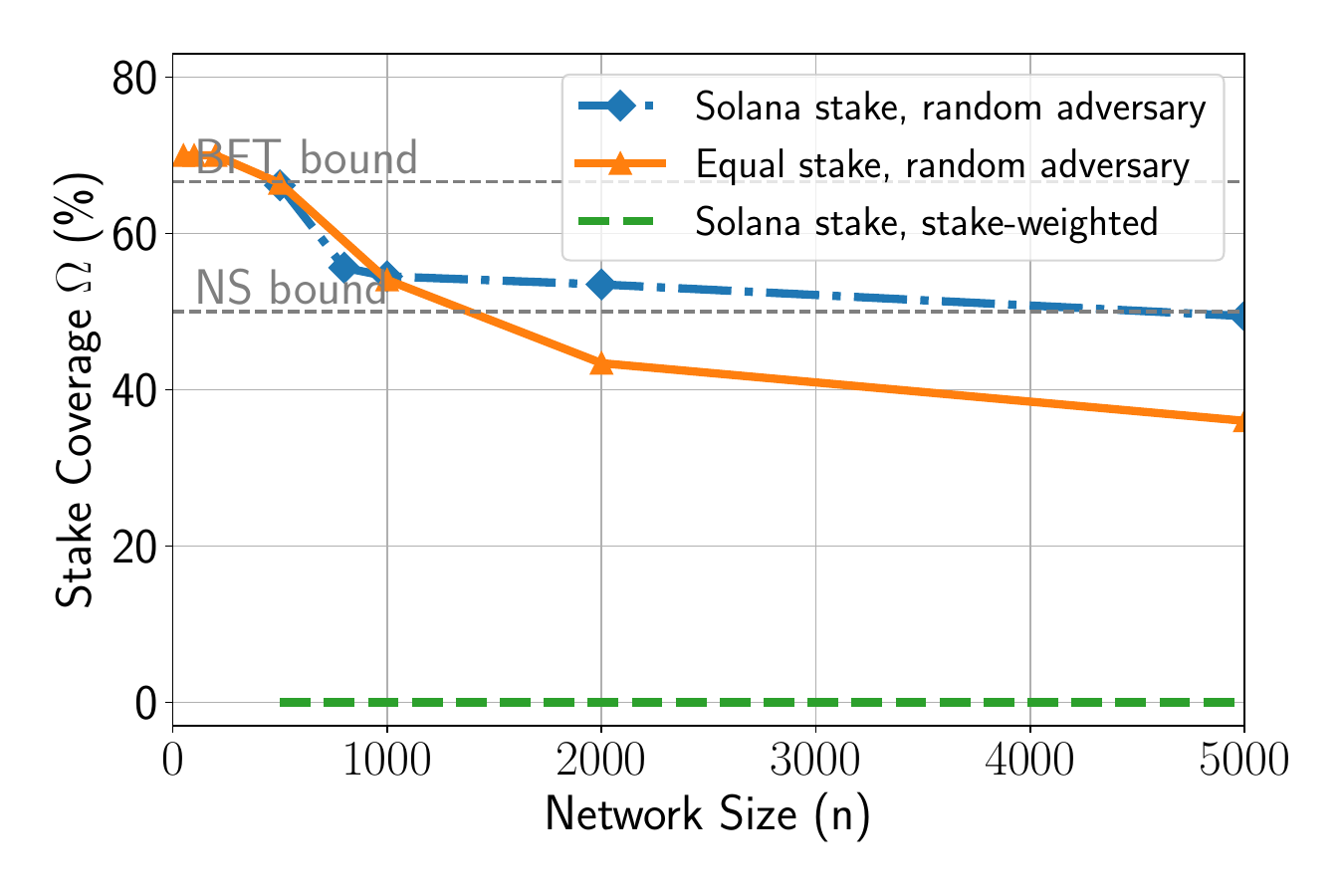}
        \caption{ Coverage in Solana under real-world stake distribution. }
        \label{fig:solana_stake_ablation}
    \end{minipage}
    \hfill
    \begin{minipage}[t]{0.52\linewidth}
        \vspace{0pt}
        \centering
        \scalebox{0.5}{
        \begin{tabular}{|c||c|c|c|c|}
            \hline
            \multirow{2}{*}{Churn rate} & \multicolumn{4}{c|}{Fraction of nodes that are online and benign }  \\
            & $\bound=0.0$ & $\bound=0.2$ & $\bound=0.3$ & $\bound=0.5$ \\
            \hline \hline
            $\churnrate =  5\%$ & \cellcolor[HTML]{B8D9F5} 98.3\% & \cellcolor[HTML]{B8D9F5} 78.6\% & \cellcolor[HTML]{B8D9F5}68.8\% & \cellcolor{gray!20}49.2\% \\ 
            $\churnrate = 10\%$ & \cellcolor[HTML]{B8D9F5} 96.8\% & \cellcolor[HTML]{B8D9F5} 77.3\% & \cellcolor[HTML]{B8D9F5}67.7\% & \cellcolor{gray!20}48.4\% \\ 
            \hdashline
            $\churnrate = 20\%$ & \cellcolor[HTML]{B8D9F5} 93.5\% & \cellcolor[HTML]{B8D9F5} 74.9\% & \cellcolor[HTML]{FDDCCC} 65.3\% & \cellcolor{gray!20}46.7\% \\ 
            $\churnrate = 30\%$ & \cellcolor[HTML]{B8D9F5} 90.0\% & \cellcolor[HTML]{B8D9F5} 72.1\% & \cellcolor[HTML]{FDDCCC} 63.1\% & \cellcolor{gray!20}44.9\% \\ 
            \hdashline
            $\churnrate = 40\%$ & \cellcolor[HTML]{B8D9F5} 86.9\% & \cellcolor[HTML]{B8D9F5} 69.3\% & \cellcolor[HTML]{FDDCCC} 60.5\% & \cellcolor{gray!20}43.6\% \\ 
            $\churnrate = 50\%$ & \cellcolor[HTML]{B8D9F5} 83.4\% & \cellcolor[HTML]{B8D9F5} 66.8\% & \cellcolor[HTML]{FDDCCC} 58.4\% & \cellcolor{gray!20}41.7\% \\
            \hline
        \end{tabular}
        }
        \captionof{table}{Impact of churn on quorum formation: BFT and NS quorums (blue), NS only (peach), neither (gray).}
        \label{tab:avg_online_stats}
    \end{minipage}
\end{figure}

\begin{figure}[h!]
    \centering
    \begin{minipage}[t]{0.49\linewidth}
        \centering
        \includegraphics[width=\linewidth]{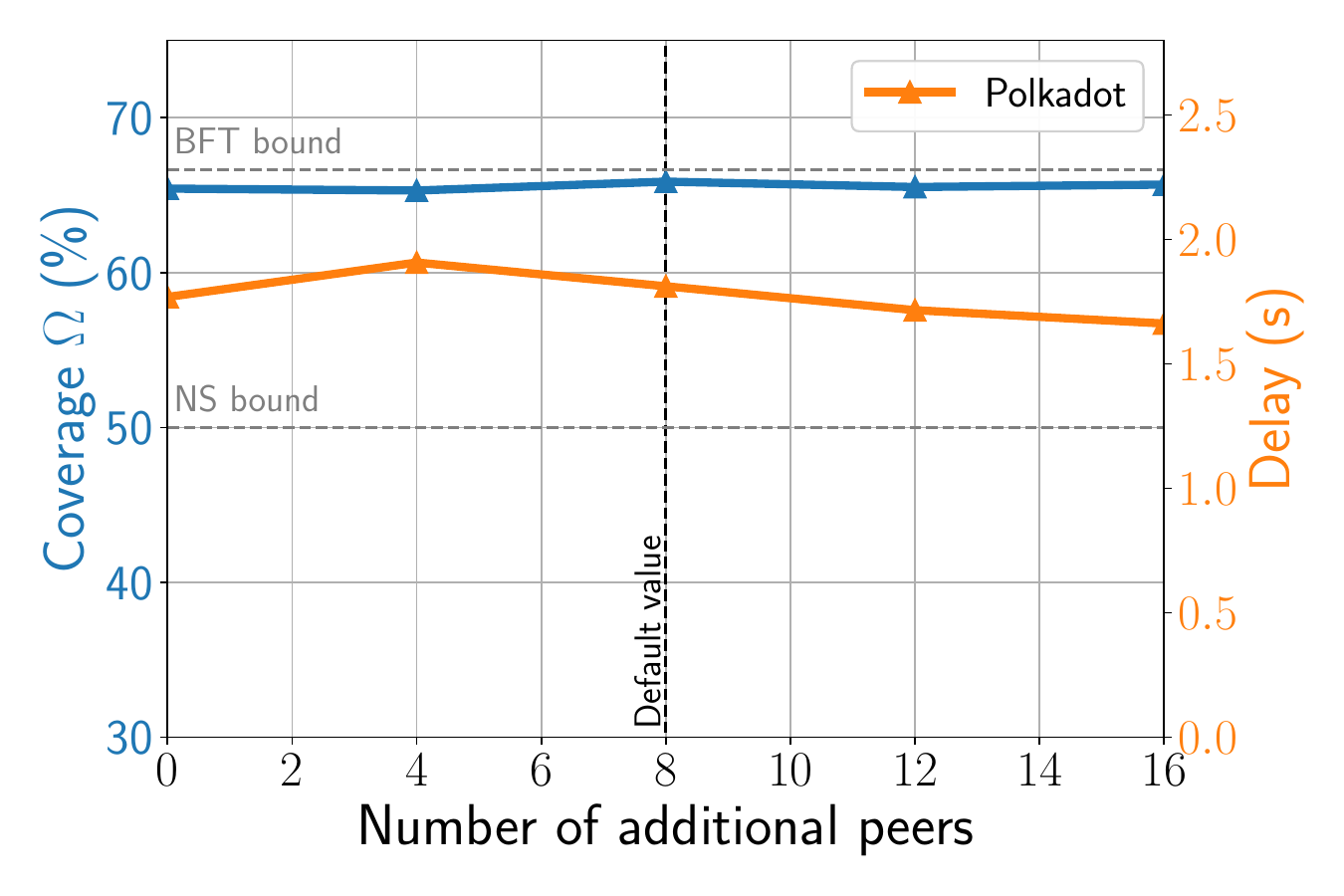}
        \caption{Impact of additional randomly selected nodes in Polkadot. }
        \label{fig:additional_nodes_ablation_Polkadot}
    \end{minipage}%
    \hfill
    \begin{minipage}[t]{0.49\linewidth}
        \centering
        \includegraphics[width=\linewidth]{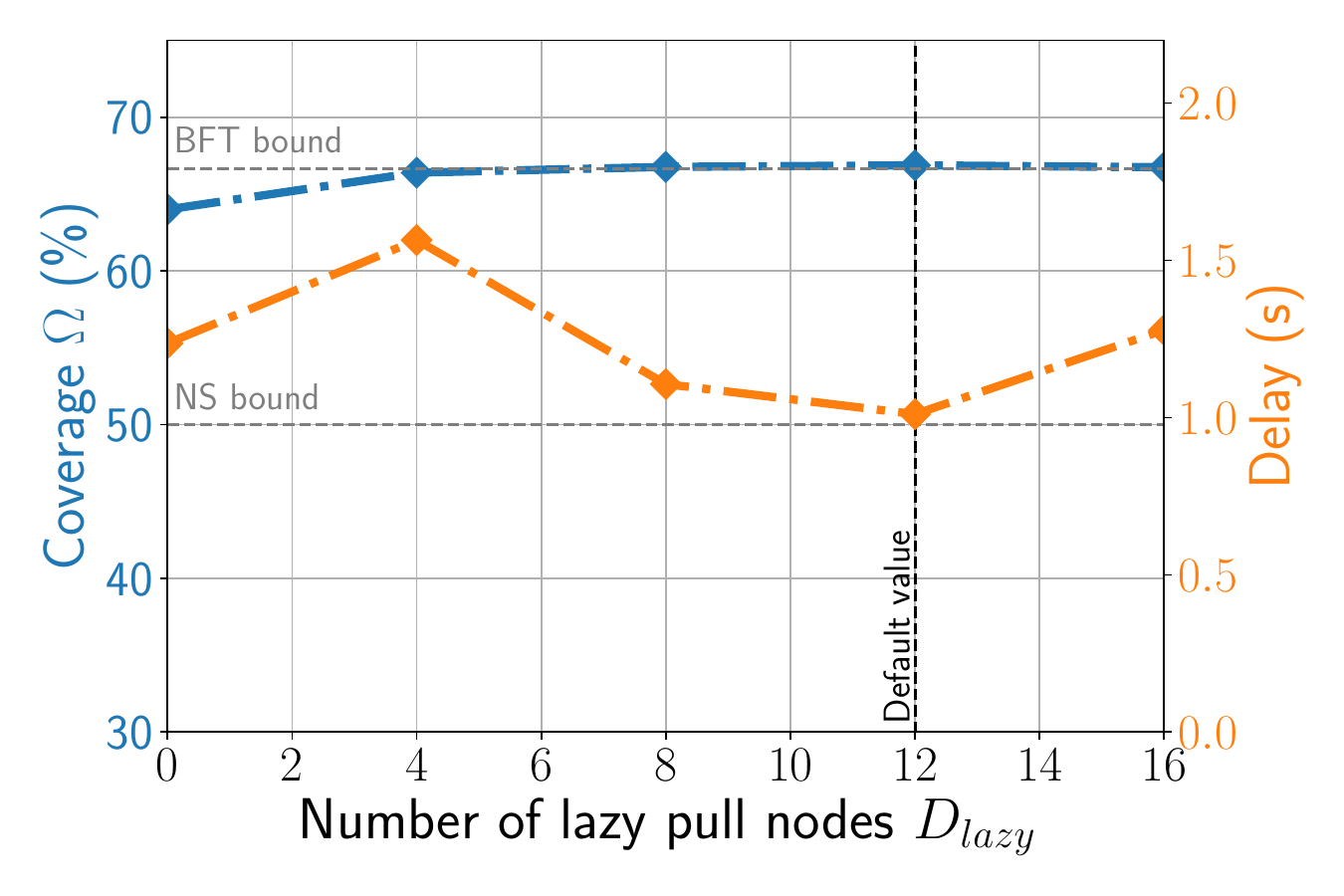}
        \caption{Impact of the number of lazy pull nodes $D_{lazy}$ in GossipSub.}
        \label{fig:additional_nodes_ablation_GossipSub}
    \end{minipage}
\end{figure}
\section{Ethics Considerations}

This work investigates the robustness of widely deployed gossip protocols under realistic adaptive adversarial strategies, with the overarching goal of advancing the security and resilience of existing decentralized platforms. While our findings indicate potential limitations in the robustness of Solana’s gossip protocol and GossipSub to adaptive attacks, all experiments were conducted exclusively within a controlled simulation environment. We did not interfere with the operation of any live Solana or GossipSub network nodes. Namely, all data used to validate our simulator (in Section~\ref{sec:methodology} and Appendix~\ref{sec:validation}) was obtained from publicly available trackers and telemetry services.

Some of our results depend on partial knowledge of the underlying network topology. Where topology information is not publicly available, we rely exclusively on methods established in prior work, rather than introducing new topology extraction techniques, a deliberate choice given the ethical considerations involved. Note that Ethereum's network topology, unlike Solana's, is not publicly disclosed, though it can be inferred through active probing. Accordingly, Figure 4 presents the Ethereum Testnet topology as reconstructed via transaction isolation by prior work, rather than the live Ethereum mainnet.

Nevertheless, we disclosed our findings to the relevant stakeholders: in particular, to Solana to highlight the lack of robustness in their gossip protocol, and to the GossipSub developers to inform them of its sensitivity to misconfigurations.

\newpage

\section{Meta-Review}

The following meta-review was prepared by the program committee for the 2027
IEEE Symposium on Security and Privacy (S\&P) as part of the review process as detailed in the call for papers.

\subsection{Summary}

This work compares the gossip and topology of several major blockchain protocols including an Erdos-Renyi random graph for Bitcoin, Polkadot's grid network, the Gossip Sub protocol used by Ethereum and a few other protocols and Solana's dynamic tree-based gossip protocol. Through simulations of several network parameters, the paper compares how much of the network receives new messages and the maximum delay to get a message to these nodes. They also present various adversarial models (random corruption and selective corruption given different observability) with the goal of delaying announcements, or not participating in gossip at all.

\subsection{Scientific Contributions}
\begin{itemize}

\item Creates a New Tool to Enable Future Science

\item Provides a Valuable Step Forward in an Established Field
\end{itemize}

\subsection{Reasons for Acceptance}
\begin{enumerate}

\item Provides a Valuable Step Forward in an Established Field: Important issue for all blockchains that is often overlooked; By covering the gossip protocols of the most important protocols in the blockchain space, the research is relevant for many real-world systems. The approach is cross-layer studying the impact of network on consensus.

\item Creates a New Tool to Enable Future Science: The authors present a new simulation framework. In-depth experimental study of the trade-offs faced with regard to gossip in terms of efficiency and robustness in the face of various adversarial models.
\end{enumerate}

\end{document}